\documentclass[prx,twocolumn,10pt,aps,superscriptaddress,printnumbers,amsmath,amssymb,showpacs,longbibliography]{revtex4-1}

\newcommand{\be}{\begin{equation}}
\newcommand{\ee}{\end{equation}}
\newcommand{\ba}{\begin{eqnarray}}
\newcommand{\ea}{\end{eqnarray}}
\usepackage{slashed}
\usepackage{graphicx}
\usepackage{dcolumn}
\usepackage{bm}
\usepackage{setspace}
\usepackage{color}
\usepackage[normalem]{ulem}

\usepackage{float}
\restylefloat{table}

\usepackage{lipsum}

\newcommand{\nh}{\bm{{\hat n}}}
\newcommand{\rv}{\bm{r}}

\newcommand{\fa}{\bm{f}_a}

\newcommand{\vEOM}{2a}
\newcommand{\nEOM}{2b}
\newcommand{\continuity}{2c}

\begin{document}

\title{The geometry of thresholdless active flow in nematic microfluidics}

\author{Richard Green}
\affiliation{Instituut-Lorentz, Universiteit Leiden, 2300 RA Leiden, The Netherlands}
\author{John Toner}
\affiliation{Department of Physics and Institute of Theoretical
Science, University of Oregon, Eugene, OR $97403^1$}
\affiliation{Max Planck Institute for the Physics of Complex Systems, N\"othnitzer Str. 38, 01187 Dresden, Germany}
\author{Vincenzo Vitelli}
\affiliation{Instituut-Lorentz, Universiteit Leiden, 2300 RA Leiden, The Netherlands}

\begin{abstract} 

``Active nematics" are orientationally ordered but apolar fluids composed of interacting 
constituents individually powered by an internal source of energy. When activity exceeds a 
system-size dependent threshold, spatially uniform active apolar fluids undergo a hydrodynamic 
instability leading to spontaneous macroscopic fluid flow. Here, we show that a special class of spatially non-uniform 
configurations of such active apolar fluids display  laminar (i.e., time-independent) flow even for 
arbitrarily small activity. We also show that two-dimensional active nematics confined on a 
surface of non-vanishing Gaussian curvature must necessarily experience a non-vanishing active 
force. This general conclusion follows from a key result of differential geometry: geodesics must 
converge or diverge on surfaces with non-zero Gaussian curvature. We derive the conditions 
under which such curvature-induced active forces generate ``thresholdless flow" for two-dimensional curved shells. We then extend our analysis to bulk systems and show how to induce 
thresholdless active flow by controlling the curvature of confining surfaces, external fields, or 
both. The  resulting laminar flow fields are  determined analytically in three experimentally 
realizable configurations that exemplify this general phenomenon: i) toroidal shells with planar 
alignment, ii) a cylinder with non-planar boundary conditions, and iii) a ``Frederiks cell" that 
functions like a pump without moving parts. Our work suggests a robust design strategy for 
active microfluidic chips and could be tested with the recently discovered ``living liquid crystals". 

\vspace{1 pc}

\noindent Subject Areas: Soft Matter

\end{abstract}

\maketitle

Active liquids \cite{Marchetti2013} are complex  fluids with some components individually capable of 
converting internal energy into sustained motion. These ``active components" can be 
sub-cellular (such as  microtubules powered by molecular motors,  and acto-myosin networks \cite{actin, microtub}),  synthetic (e.g., self-propelled colloids \cite{Janus}, or interacting micro-robots), or, alternatively, living organisms \cite{boids1, boids2, boids3, boids4}, such as birds, fish \cite{couzin}, microorganisms \cite{dictyo1, dictyo2} or insects \cite{midges}. Hybrid systems composed of motile rod-shaped bacteria placed in nontoxic liquid crystals have also been recently realized \cite{living}. All of these systems blur the line between the living and synthetic world, thereby  opening up unprecedented opportunities for the design of novel smart materials and technology. At the same time, the far-from-equilibrium nature of active matter leads to exotic phenomena of fundamental interest. Among these are the ability of active fluids to (i) spontaneously break a continuous symmetry in two spatial dimensions \cite{TT1,TT2,TT3,TT4}, (ii) exhibit spontaneous steady state flow \cite{actnem, actin} in the absence of an external driving force and (iii) support topologically protected excitations (e.g., sound modes) that originate from time-reversal symmetry breaking \cite{Souslov2016}.

A striking example of the phenomenon of spontaneous flow occurs in active nematic liquid crystals \cite{Marchetti2013, actnem, Keber2014, Giomi2013, Giomi2014, Sanchez2012, Ravnik2013}. These materials are orientationally ordered but apolar fluids; that is, the active particles share a common axis of motion but, in the homogeneous state, equal numbers of them move in each of the two directions parallel to this axis. As a result, there is no net motion and no net flow. However, if the activity parameter $\alpha$ (defined later) exceeds a {\it critical  threshold} $\alpha_c$, the undistorted nematic ground state becomes unstable. Once this instability threshold is passed, the active nematics spontaneously deform their state of alignment, triggering macroscopic ``turbulent flow" \cite{actnem, actin, Marenduzzo2007, GiomiPRL2008, Giomi2012, Edwards2009}.
For nematics, this activity threshold $\alpha_c$ goes to zero  as the system size $L\rightarrow\infty$: $\alpha_c\sim {K\over L^2}$, where $K$ is a characteristic Frank elastic constant.
Equivalently, one can say that  the instability-triggered flow does not occur in  systems of characteristic size smaller than $ L_{\text{inst}}\sim\sqrt{\frac{K}{|\alpha|}}$.
 
Numerical studies of active nematics suggest that some non-uniform director configurations can
lead to {\it laminar} flow for arbitrarily small activity, i.e., well below the instability threshold \cite{Marenduzzo2007}. However, no systematic study of 
the mechanisms and criteria behind such ``thresholdless active flow" has previously been undertaken. In this paper, we use a well established hydrodynamic theory of active nematics to identify the class of surface deformations, boundary conditions or external fields that induce a non-uniform director ground-state capable of generating such thresholdless laminar flow. We emphasize that not all spatially non-uniform configurations will induce such flow.

The condition for a given set of boundary conditions and applied fields to induce thresholdless active flow in nematics is most easily expressed in terms of the director field $\nh(\rv)$ \cite{deGennes}, which is defined as the local orientation of molecular alignment. It can be stated as follows: if the active force, which is defined in general as 
\begin{equation}
{\bf f}_a\equiv \alpha\left[\bm{{\hat n}}\left(\nabla \cdot \bm{{\hat n}}\right) +(\bm{\hat{n}} \cdot \nabla) \bm{\hat{n}} \right],
\end{equation}
has non-zero curl, when computed for the director configuration $\bm{{\hat n}}({\bf r})$ that minimizes the Frank elastic free energy {(including external fields)} of the corresponding equilibrium problem \cite{deGennes}, then the active fluid in the same 
geometry {\it must} flow (i.e., the velocity field ${\bf v}\ne{\bf 0}$). Note that this condition is far more stringent than simply requiring that the nematic ground state orientation be inhomogeneous. For example,  {\it any} pure twist configuration (e.g, a cholesteric, or a twist cell) does not satisfy it, since splay $\nabla \cdot \bm{{\hat n}}$ and bend $ \hat{\bm{n}} \times \left( \nabla \times \hat{\bm{n}} \right)$ both vanish in such configurations. The criterion ${\bf \nabla \times f}_a \ne {\bf 0}$ is a sufficient but not necessary condition for thresholdless flow.

In the present study, we calculate the resulting flow field ${\bf v}({\bf r})$ explicitly in the ``frozen director" approximation, in which 
the nematic director remains in its equilibrium configuration when activity is turned on. We demonstrate that this approximation is asymptotically exact in the experimentally relevant limit of weak orientational order. Since many nematic to isotropic transitions are weakly first order \cite{deGennes} (at least in equilibrium), this frozen director limit may be realized close to such transitions. Moreover, these approximate solutions provide qualitative insights into the nature of the flow.
 
Our ideas can also be applied with minor modifications to the recently discovered ``living liquid crystals" \cite{living}. These systems are a {\it mixture} of two components: living bacteria, which provide the activity, and a background medium composed of nematically ordered non-active molecules. On symmetry grounds, as we discuss in more detail  in the Supplemental Material \cite{SM1} section I, these systems considered in their entirety {\it are} active nematics.

The  remainder of this paper
is organized as follows:
in section I, we 
review the ``standard model" for the hydrodynamics of active nematics.
We also discuss some generalizations of this model, and argue that none of our conclusions will be substantively affected by these generalizations.
In section II, we derive the general criterion for thresholdless active flow.
In section III, we apply this criterion to the specific case of surfaces of non-zero Gaussian curvature, and show that such surfaces {\it always} have non-zero active forces, but need not always have thresholdless flow. We also derive the additional criteria that must be satisfied for thresholdless flow to occur in these systems,  give a specific example (active nematics on a thin toroidal shell) in which these conditions are met, and work out the flow field in this case. In section IV, we derive similar results for {\it bulk} systems with curved boundaries. Section V presents calculations, in the frozen director approximation, of the flow fields that result in microchannels with prescribed anchoring angles on the surface, and section VI presents a novel design for an
``active pump" in which a Frederiks cell geometry is used to ``switch on" thresholdless active flow.

\section{The hydrodynamics of active nematics}
We take as our model for an {\it incompressible}, one-component active nematic fluid  the following three coupled equations \cite{actnem}:
\begin{subequations}
\begin{eqnarray}
\rho_0 \frac{D v_k}{D t} &=& - \partial_k P + \eta \nabla^2 v_k + \alpha \partial_j \left(n_j n_k  \right)  
+\partial_j(\lambda_{ijk}{\delta F\over\delta n_i})
 \nonumber \\
  \label{eq: vEOM} \\ 
\frac{D n_i}{D t}
&=& \lambda_{ijk} \partial_jv_k-
 \frac{1}{\gamma_1} \left[{\delta F\over \delta n_i} - \left({\delta F\over\delta\hat{n}} \cdot \hat{n} \right) n_i\right]
\label{eq: nEOM} \\
\nabla \cdot \bm{v} &=& 0, \label{eq: continuity}
 \end{eqnarray}
 \end{subequations}
 
\noindent where $D/Dt \equiv \partial_t + \bm{v} \cdot \nabla$ is the convective derivative and the tensor $\lambda_{ijk}$ reads
\begin{equation}
\lambda_{ijk}\equiv\left({\lambda+1\over 2}\right)n_j\delta_{ik}+\left({\lambda-1\over 2}\right)n_k\delta_{ij}-\lambda n_in_jn_k.
\label{lambdadef}
\end{equation}
 
The first Eq. \noindent (\vEOM) is a modified Navier-Stokes equation describing the evolution of the velocity field $\bm{v}(\bm{r}, t)$; Eq. (\nEOM) is the nematodynamic equation describing the evolution of the director field $ \hat{\bm{n}}(\bm{r}, t)$, which responds both to the flow $\bm{v}$, and to its own molecular field $\frac{\delta F}{\delta \bm{n}}$ (described in more detail below), and (\continuity) is the incompressibility condition, which is required since we take the density $\rho_0$ to be constant. We denote by $P$ the dynamic pressure, $\eta$ the shear viscosity, which we take to be isotropic for simplicity, and $\gamma_1$ the director field rotational viscosity. The dimensionless flow-alignment parameter $\lambda$ captures the anisotropic response of the nematogens to shear. Note that the only difference between Eq. (\vEOM-\continuity) and the equations of motion for an equilibrium nematic \cite{deGennes} is the active force term $\alpha \partial_j \left(n_j n_k  \right)$ in the Navier-Stokes Eq. (\vEOM), which may be contractile ($\alpha>0$) or extensile ($\alpha<0$), depending on the system \cite{actnem}. The molecular field $\frac{\delta F}{\delta \bm{n}}$, derived from the  Frank free energy 
\begin{eqnarray} 
F &=& \frac{1}{2}\int d^3r[K_{1} \left(\nabla\cdot\hat{\bm{n}}\right)^2+ K_{2}\left(\hat{\bm{n}}\cdot \left( \nabla \times \hat{\bm{n}} \right)\right)^2 \nonumber \\
&& \>\>\>\>\>\>\>\>\>\>\>\>\>\>\>\> + K_{3}\left|\hat{\bm{n}}\times \left( \nabla \times \hat{\bm{n}} \right)\right|^2 ]\,,
\label{Frank FE}
\end{eqnarray}

\noindent is parametrized respectively by three independent elastic constants $K_{1,2,3}$  for  splay, twist, and bend deformations of the director. 

Many experimental realizations of active nematics, such as the ``living liquid crystals" \cite{living}  are {\it multi-component} systems. This leads to some differences between their hydrodynamic theories and that embodied by Eq. (\ref{eq: vEOM}-\ref{eq: continuity}); in particular, the concentration of each additional component must be added as a new hydrodynamic variable. However, the resulting hydrodynamic equations are sufficiently similar to the one component case we consider here that our criteria for thresholdless flow remain valid; we will demonstrate this in a future publication. 
We discuss the applicability of a multi-component generalization of the hydrodynamic equations presented here to living liquid crystals in the Supplemental Material \cite{SM1} section I.

We also note that, strictly speaking, Eq. (\ref{eq: vEOM}-\ref{eq: continuity})
are {\it not} the most general set of equations for a one-component active nematic. Specifically, there are two ways  in which they could be generalized:

\noindent1) The free energy $F$ that appears in the velocity equation of motion (\ref{eq: vEOM}) need not, in a non-equilibrium system, be the same as that in the director equation of motion (\ref{eq: nEOM}). Both free energies have to have the same {\it form} as (\ref{Frank FE}), since that form is required by rotation invariance, but the Frank constants $K_{1,2,3}$ that appear in them need not be equal.

\noindent2) The viscosity need not be isotropic: there are in general six ``Leslie coefficients" \cite{Leslie} characterizing this anisotropic response.

In the Supplemental Material \cite{SM1} section II, we show that, for small activity, the former effect is negligible: the Frank constants in the velocity equation of motion (\ref{eq: vEOM}) approach those in the director equation of motion (\ref{eq: nEOM}) as activity goes to zero. Hence, since we are interested here in the limit of small activity, this difference becomes negligible. Furthermore, for weak orientational order, our approximation of isotropic viscosity becomes asymptotically exact. Indeed,  Kuzuu and Doi \cite{Doi}  show that the anisotropic parts of the viscosities vanish as the amplitude of the nematic order parameter goes to zero (the isotropic viscosity does not vanish in the same limit since a completely isotropic liquid has a non-vanishing isotropic viscosity). To sum up, the hydrodynamic theory we employ is exact for low activity, weakly ordered nematics. 

Most of our conclusions are independent of this limit; in particular, our criteria for thresholdless active flow are dictated only by form of active force ${\bf f}_a$, which is unchanged by making viscosity tensor anisotropic, and which is furthermore independent of the assumption that the two Frank energies $F_n$ and $F_v$ are the same.

\section{Thresholdless flow in active nematics}
 
At the heart of our study lies a simple observation: the constitutive Eq. (\ref{eq: vEOM}-\ref{eq: continuity}) of active nematics inevitably imply that, in certain geometries, an arbitrarily small activity  induces steady state macroscopic fluid flow. We will prove this by contradiction. If there is {\it no} fluid flow (i.e., if the velocity field ${\bf v}={\bf 0}$), then the  equation of motion (\nEOM) for the director field implies that, in a steady state, for which $\frac{D n_i}{D t}=0$, $\frac{\delta F}{\delta \bm{n}} - (\bm{{\hat n}} \cdot \frac{\delta F}{\delta \bm{n}}) \bm{{\hat n}}={\bf 0}$ (which also holds in the case of anisotropic viscosity). This  is simply the Euler-Lagrange equation for minimizing the Frank free energy $F$ subject to the constraint $|\nh|=1$. The contradiction arises when we insert such an equilibrium solution for the nematic director into the 
equation of motion for the velocity field (\vEOM).

The last term on the right hand side of Eq. (\vEOM), involving $\frac{\delta F}{\delta \bm{n}}$, vanishes when $\frac{\delta F}{\delta \bm{n}} \parallel \bm{\hat{n}}$, which is the case when the director field is in its ground state. Since the velocity field ${\bf v}$ vanishes, Eq. (\vEOM)
reduces to $ \nabla P =\alpha \left( \bm{{\hat n}} \cdot \nabla \bm{{\hat n}} + \bm{{\hat n}}\nabla \cdot \bm{{\hat n}}\right)\equiv\fa$. Hence the pressure gradient must cancel the active force to prevent flow, but this is not possible if the active 
force has a non-vanishing curl. In such cases, ${\bf v}={\bf 0}$ can {\it never} be a 
solution in the presence of activity; the fluid {\it must flow}, no matter how small the activity. Thus, a sufficient (but not necessary) condition for thresholdless active flow is
\begin{equation}
\nabla \times \bm{f}_a \ne \bm{0},
\label{condition}
\end{equation}
\noindent which has also been implicit in other work such as \cite{Thampi2014}.

One class of director configurations for which the condition in Eq. (\ref{condition}) is {\it not} satisfied is that of ``pure twist" configurations; that is, configurations in which the twist does not vanish (i.e., $\hat{\bm{n}} \cdot \left( \nabla \times \hat{\bm{n}}\right) \ne 0$), but the splay and bend do (i.e.,  $ \nabla \cdot \hat{\bm{n}}=0$ and  $\hat{\bm{n}} \times \left( \nabla \times \hat{\bm{n}}\right)=\bm{0}$, respectively). This can be seen by using the vector calculus identity $[\hat{\bm{n}} \times \left( \nabla \times \hat{\bm{n}}\right)]_{_i}=n_j\nabla_in_j-\bm{{\hat n}} \cdot \nabla n_i={1\over 2}\nabla_i|\hat{n}|^2-\bm{{\hat n}} \cdot \nabla n_i=-\bm{{\hat n}} \cdot \nabla n_i$, where in the last equality we have used the fact that $\hat{\bm{n}}$ is a unit vector to set $\nabla_i|\hat{n}|^2=\nabla_i 1=0$. Using this, the active force $\bm{f}_a$ may be rewritten as
\begin{equation}
\label{eq: active force}
\bm{f}_a = \alpha \left[ \hat{\bm{n}} \> \nabla \cdot \hat{\bm{n}} - \hat{\bm{n}} \times \left( \nabla \times \hat{\bm{n}} \right) \right],
\end{equation}
\noindent which implies that a director field with pure twist has zero active force, and, hence, no flow for sufficiently small activity. 

When we consider specific examples of thresholdless active flow in the remainder of this paper, we will determine analytically the velocity field $\bm{v}(\bm{r}, t)$. In general, this is a difficult, non-linear calculation, since the flow field reorients the nematic director. However, in the ``frozen director" limit $\gamma_1\ll\eta$,  turning on activity (and thereby inducing thresholdless flow) does not lead to an appreciable change in the nematic director configuration from that in equilibrium, which is obtained by minimizing the Frank free energy. We show in the Supplemental Material \cite{SM1}, section III that there is a very natural, generic, and well-defined limit in which $\gamma_1$ will always be much less than $\eta$: namely, the limit of weak nematic order. 

\begin{figure*}
\centering
\includegraphics[width=.8\textwidth]{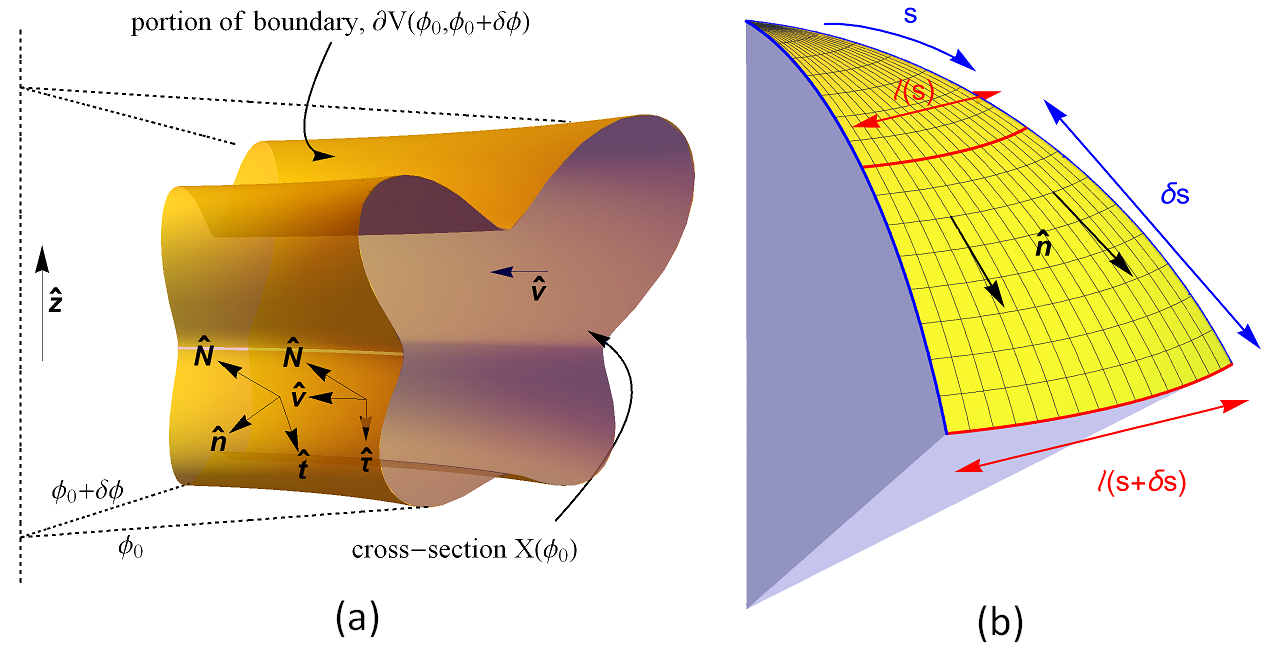}
 \caption{Color) (a) A volume $V'$ of arbitrary cross-section with torsional symmetry. The normal to the bounding surface is $\bm{\hat{N}}$ and 2 orthonormal sets of unit vectors are shown: (i) director field $\bm{\hat{n}}$ tangential to the bounding surface with $\bm{\hat{t}}=\bm{\hat{N}} \times \bm{\hat{n}}$; and (ii) direction of symmetry $\bm{\hat{\nu}}$ also tangential to bounding surface with $\bm{\hat{\tau}}=\bm{\hat{N}} \times \bm{\hat{\nu}}$; and (b) In the case of planar anchoring of the director $\bm{\hat{n}}$ on a surface with Gaussian curvature, the distance between geodesics $\ell(s)$ as a function of the arc-length $s$. }
\label{fig: pillbox}
\end{figure*}

\section{\label{Sec: curved}Two-dimensional curved systems}
\subsection{\label{Sec: curvedgen}General Considerations}

Consider an active nematic material confined to a curved monolayer shell, such as that shown in Fig. \ref{fig: pillbox}. Such systems are of special interest since many active nematics synthesized to date are monolayers or thin shells with planar anchoring \cite{Keber2014, Sanchez2012}. In this section, we will show that, in general, a shell with non-vanishing Gaussian curvature $G$ generates a non-vanishing active force $\bm{f}_a$. To prove this result, we first assume that, if the shell is very thin, the component of $\bm{\hat{n}}$ perpendicular to the surface is negligible everywhere inside the shell \cite{Vitelli2006, Lopez-Leon2011, Fernandez-Nieves2007}, i.e., planar anchoring conditions. In this case, we can decompose the active force $\bm{{f}}_a(\bm{x})$ at position $\bm{x}$ along three orthogonal directions: (i) the local surface normal $\bm{{\hat N}}$, (ii) the nematic director $\bm{{\hat n}}$ and (iii) the tangent vector $\bm{{\hat t}}$ perpendicular to both $\bm{{\hat N}}$ and $\bm{{\hat n}}$, shown in Fig.  \ref{fig: pillbox}(a) (which in addition shows a second orthonormal set of unit vectors $(\bm{{\hat N}}, \bm{\hat{\nu}}, \bm{\hat{\tau}})$ used below in Section IV). The active force reads

\begin{eqnarray}
\label{eqn:f}
\bm{{f}}_a(\bm{x})= \alpha \left[ \bm{{\hat n}}(\bm{x})  \!\!\!\! \quad \nabla \cdot \bm{{\hat n}}(\bm{x}) + \bm{{\hat t}}(\bm{x}) \!\!\!\! \quad \kappa_g(\bm{x}) + \bm{{\hat N}}(\bm{x}) \!\!\!\! \quad \kappa_n(\bm{x}) \right] \nonumber \\
\label{eq: f}
 \end{eqnarray}

\noindent where $\kappa_n = \bm{{\hat N}} \cdot  \left( \bm{{\hat n}} \cdot \nabla \right) \bm{{\hat n}}$ denotes the local normal curvature of the nematic director field $\bm{{\hat n}}(\bm{x})$  and $\kappa_g = \bm{{\hat t}} \cdot  \left( \bm{{\hat n}} \cdot \nabla \right) \bm{{\hat n}}$ denotes its geodesic curvature \cite{RevModPhys.74.953, Santangelo2007}, which quantifies deviations from the local geodesic tangent to $\bm{\hat{n}}$. 

Since the set of vectors $(\bm{{\hat N}}, \bm{{\hat n}},\bm{{\hat t}})$ is orthonormal, the active force can only vanish if {\it all three} of its components vanish. In particular, this implies that $\kappa_g=0$. However, we now show that the condition $\kappa_g=0$ forces the $\bm{{\hat n}}$ component of $\bm{{f}}_a$ (which is proportional to $ \nabla \cdot \bm{{\hat n}}$) to be non-zero, on any surface with non-zero Gaussian curvature. To prove this statement, note that if $\kappa_g=0$, the nematic director must lie on geodesics everywhere on the surface, as illustrated in Fig.  \ref{fig: pillbox}(b). Consider an \emph{infinitesimal} patch bounded by two geodesics (along which the nematic director is aligned) and their normals, drawn in red in Fig.  \ref{fig: pillbox}(b). These perpendicular ``arcs" have length equal to the distance $\ell(s)$ between the two geodesics parametrized by the arc-length $s$ along one of them. We now apply the divergence theorem to the director field $ \bm{\hat{n}}$ on this small patch, whose area is approximately given by $ds$ times $\ell (s)$. The $\bm{\hat{n}}$ flux vanishes along the two geodesics, and it is equal to $\ell (s+ds)$ and $-\ell (s)$ along the two red arcs, which yields
\begin{equation}
\nabla \cdot \bm{{\hat n}} = \frac{1}{\ell} \frac{d \ell}{d s}.
\label{central2}
 \end{equation}

The right hand side of Eq. (\ref{central2}) cannot be identically zero because $\frac{d^2 \ell}{ds^2}=-G(s) \> \ell$ on an arbitrary surface with non-vanishing $G(x)$ \cite{difgeo}. Intuitively, Gaussian curvature forces geodesics to either converge or diverge,  which in turn implies that $\nabla \cdot \bm{\hat{n}}\ne 0$. The converse statement also holds, namely that $\nabla \cdot \bm{\hat{n}}= 0$ requires $\kappa_g \ne 0$. Thus we have proved that non-vanishing Gaussian curvature $G$ implies a non-vanishing active force $\bm{f}_a$. The incompatibility relation derived above has a purely geometric origin and it is also responsible for the geometric frustration of nematic (and more generally orientational and crystalline) order in curved space.
This general result is independent of specific choices of elastic constants and other material parameters, such as the viscosity tensor. 

A non-vanishing Gaussian curvature always enforces a non-zero in-plane active force, but thresholdless flow will occur only if this active force $\bm{f}_a$ cannot be balanced by the pressure gradient. Since $\nabla P$ is by definition a conservative force, a sufficient condition for thresholdless flow is therefore
\begin{equation}
G(\bm{x}) \ne 0 
\end{equation}
\noindent at some point $\bm{x}$ on the shell, and
\begin{equation}
\oint_{C} d\bm{l}\cdot \bm{f}_a \ne 0
\label{eq: cond closed loop}
\end{equation}

\noindent for some closed loop $C$ on the shell. 

Our derivation of this condition never assumed  that the director configuration was free of topological defects (i.e., disclinations); hence the active force {\it must} be non-zero for {\it any} surface with non-vanishing Gaussian curvature, even if, as often happens \cite{Vitelli2004, Bowick2004}, that Gaussian curvature induces disclinations on the surface. Indeed, topological defects, far from {\it preventing} flow, actually make it inevitable (a result first noted in references \cite{Giomi2013, Giomi2014} for flat surfaces), since they induce large director gradients near their core.

Note, however, the condition (\ref{eq: cond closed loop}) will {\it not} be satisfied for all surfaces with non-zero Gaussian curvature, even though the active force must be non-zero for all such surfaces. In the next section, we consider a specific example that illustrates this point.

\subsection{\label{Sec: torus} Chiral symmetry breaking and flow in toroidal shells}

Let us focus our analysis on the case of a curved nematic monolayer with the molecules aligned tangent to the surface of a torus, but free to choose their local in-plane orientation. Since the torus is a surface of non-zero Gaussian curvature, there must be a non-vanishing active force based on our previous reasoning. We now demonstrate that, as the aspect ratio of the torus is changed, this active force results in no net flow for very slender tori (which are nearly cylinders), while for ``fatter" tori, there is a transition to a chiral director configuration in which the active force {\it does} have a non-zero line integral. Hence, by the criteria of the previous subsection, flow must ensue. We compute such flow in the aforementioned ``frozen director" approximation.

\begin{figure}
 \centering
\includegraphics[width=.48\textwidth]{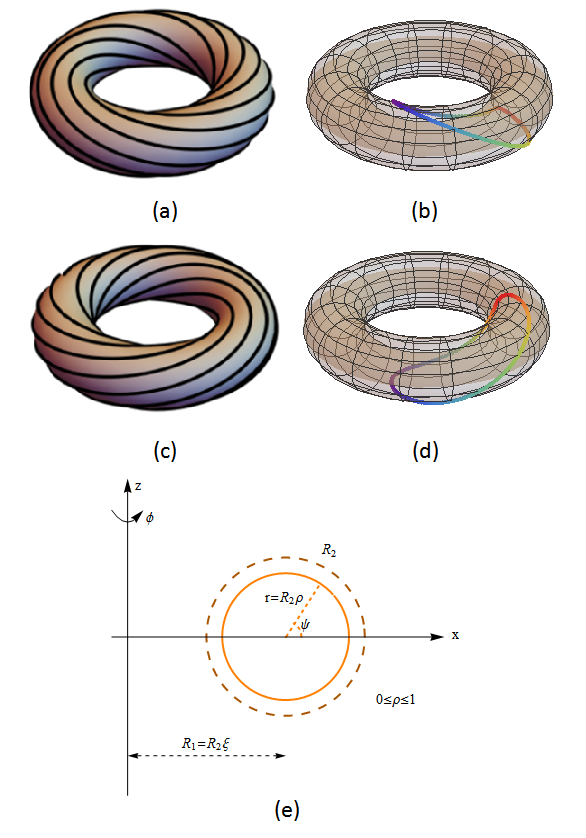}
 \caption{Color) (a) Left chiral ground state ($\omega<0$) of an active nematic liquid and associated flow on the surface (b). (c) Right chiral ground state ($\omega>0$) of an active nematic liquid and associated flow on the surface (d). Activity is contractile ($\alpha>0$); red denotes flow in the positive $\phi$-direction, violet denotes flow in the negative $\phi$-direction and green denotes no flow. (e) \emph{Toroidal co-ordinates} The plane $y=0$ is shown for the $x\ge 0$ half of the toroid using $(\rho,\psi,\phi)$ coordinates where $\psi$ is the poloidal angle and $\phi$ the azimuthal angle centered at the point $x=R_1, z=0$.}
\label{fig: Toroid combined 2D}
\end{figure}

Consider the set of toroidal coordinates $(\rho, \psi, \phi)$ shown in Fig.  \ref{fig: Toroid combined 2D}(e), where $\rho$ is the dimensionless radial coordinate set to $1$ on the monolayer surface, $\psi$ is the poloidal angle, and $\phi$ is the toroidal or azimuthal angle. 
The slenderness of the torus $\xi \equiv R_1/R_2$ is the aspect ratio of its major ($R_1$) and minor ($R_2$) radii. For very slender tori (which are nearly cylinders), the nematic director will be everywhere oriented along $\bm{\hat{\phi}}$. This bend-only configuration is divergenceless, hence the first term of Eq. (\ref{eq: f}) is zero. Note however that $\kappa_g$ will be different from zero because the nematic director lines are not geodesics. In fact, $\kappa_g = \sin \psi / R_2( \xi+ \cos \psi)$ and $\kappa_n = -\cos \psi / R_2( \xi+ \cos \psi)$, so that in this case $\bm{f}_a=-\alpha \nabla \log (\xi+\rho \cos \psi)$. Condition (\ref{eq: cond closed loop}) is not satisfied, and there is no flow because the active force is completely balanced by the pressure gradient.

For sufficiently ``fat" (i.e., large $\xi$) tori, this uniform azimuthal director state becomes unstable to one which has non-zero twist. We extend the approach used in \cite{Pairam2013, Koning2014, Davidson2015} to a two-dimensional curved monolayer by considering the following variational ansatz which captures the qualitative features of the chiral symmetry breaking transition in the ground state at zero activity:

\begin{equation}
\bm{{\hat n}}=\frac{\omega \xi}{\xi + \cos \psi} \>\> \hat{\bm{\psi}}+\sqrt{1-\left(\frac{\omega \xi}{\xi + \cos \psi}\right)^2} \>\> \hat{\bm{\phi}},
\label{equation: ansatz}
\end{equation}
\noindent where $\omega$ is a variational parameter describing the degree of twist in the director field.

In the Supplemental Material \cite{SM1} section VII, we show that to leading order in $1/\xi$, the two ground states in Fig. \ref{fig: Toroid combined 2D}(a) correspond to 
\begin{equation}
\omega = \pm \sqrt{\frac{5}{4 \xi^2}-\frac{K_2}{2 K_3}}, 
\end{equation}
\noindent provided that the quantity under the square root is positive (otherwise the ground state is the untwisted state $\omega=0$). To $\mathcal{O}(\omega)$, the corresponding active force reads:
\begin{eqnarray}
\bm{f}_a 
= \frac{-\alpha \left(\hat{\bm{\rho}} \cos \psi -\hat{\bm{\psi}} \sin \psi + \hat{\bm{\phi}} \frac{\omega \xi \sin \psi}{\xi + \cos \psi} \right)}{R_2 (\xi + \cos \psi)},
\label{eq: active force toroid}
\end{eqnarray}
\noindent evaluated at $\rho=1$, i.e., on the surface of the torus. Condition (\ref{eq: cond closed loop}) is now satisfied by a closed loop $C$ everywhere in the $\bm{\hat{\phi}}$-direction, and so we conclude that there must be thresholdless flow.
In a two-dimensional nematic shell draped on a substrate, momentum is not generally conserved; therefore, a frictional term $-\gamma v_k$ must be added to the right hand side of Eq. (\ref{eq: vEOM}). In the limit $\gamma \gg \eta / L^2$, where $L$ is the size of the sample, this frictional drag dominates the viscous forces, and the revised form of Eq. (\ref{eq: vEOM}) reads
\begin{equation}
\nabla P + \gamma \bm{v} = \bm{f}_a.
\label{2dvEOM}
\end{equation}
Taking the divergence of Eq. (\ref{eq: active force toroid}), we see that Eq. (\ref{2dvEOM}) can be solved to obtain the pressure  $P= -\alpha \log (\xi + \rho \cos \psi)$. Therefore, on the shell where $\rho=1$, the velocity is given by
\begin{equation}
\bm{v} = \frac{-\alpha}{\gamma R_2}\frac{\omega \xi \sin \psi}{(\xi+\cos\psi)^2} \hat{\bm{\phi}} + \mathcal{O}(\omega^2). 
\label{eqn: solution}
\end{equation}
This solution is a flow one way in the $\hat{\bm{\phi}}$-direction on the top half of the toroidal shell and in the opposite direction on the bottom half. The orientation of the flow is determined by the sign of the activity (contractile or extensile) and the chirality of the ground state, as illustrated in Fig.  \ref{fig: Toroid combined 2D}.


In the above analysis, we have used a smooth, defect-free toroidal ansatz (\ref{equation: ansatz}),  which explicitly excludes the possibility of topological defects in the nematic director configuration. This is valid in the limit of high slenderness $\xi$, in which the Gaussian curvature is too small to induce defects \cite{Vitelli2004}. However, as we have noted above, defects make flow inevitable, since they induce large director gradients near themselves. Therefore, by excluding such defects, we have actually maximized the chance of having no thresholdless flow. We therefore conclude  that for tori fatter than $\xi<\xi_c= \sqrt{\frac{5K_3}{2 K_2}}$ (the last equality holding approximately when $K_3 \gg K_2$), thresholdless active flow will {\it definitely} occur.  If disclinations are not generated, and the aforementioned conditions for the validity of the frozen director approximation hold, then the flow field should be approximately described by our result (\ref{eqn: solution}).

In Appendix B, we solve the 3D bulk version of this toroidal system with no-slip boundary conditions. This is a more complicated calculation, as Eq. (\ref{2dvEOM}) becomes $\nabla P - \eta \nabla^2 \bm{v} = \bm{f}_a$, which must be solved in the bulk toroidal geometry, but the general features of the solution are similar to the shell case.

\section{Three-dimensional systems with curved boundaries}

We now turn to investigate how the geometry of the boundaries and anchoring conditions of the director can also force thresholdless flow in bulk active nematics under confinement. This may be of practical importance, since controlling boundaries and boundary conditions for liquid crystals is a highly developed technology, that has long been used for, inter alia, the construction of liquid crystal displays. Efforts are under way to extend such control to the active regime \cite{living, Sokolov2015, Guo2016}.

Consider non-planar alignment of the director to the walls of a three-dimensional channel with torsional symmetry (by which we mean equivalently that the sample is bounded by a surface of revolution about the $z-$axis as shown in Fig.  \ref{fig: pillbox}(a)). The nematic liquid crystal fills the bulk bound by the surface. If we make the additional assumption that the pressure gradient vanishes along the direction of torsional symmetry, which we denote by $\bm{\hat{\nu}}$, a non-zero component of the active force along $\bm{\hat{\nu}}$ will result in thresholdless flow.

A small section of a channel $V'$ bounded by an arbitrarily shaped surface with torsional symmetry along $\bm{\hat{\nu}}$ is shown in Fig. \ref{fig: pillbox}(a), where the local surface normal is represented by the unit vector ${\bf \hat{N}}(\bm{x})$. Denoting the torsional coordinate by $\phi$, the volume $V'$ is the section of the three-dimensional channel bounded by the surfaces $\phi=\phi_0$ and $\phi=\phi_0+\delta\phi$. The {\it integrated} force $\bf{F}(\phi_0)$ acting on the volume $V'$ can then be obtained by integrating the force density, $(f_a)_i= \alpha \> \partial_j (n_i n_j)$, over the infinitesimal volume $V'$. Applying the divergence theorem, we obtain the projection of $\bf{F}(\phi_0)$ along $\bm{\hat{\nu}}(\phi_0)$ in terms of the anchoring conditions of the nematic director at the boundary, leading to the sufficient condition for thresholdless flow:

\begin{eqnarray}
0 \ne \bm{F}(\phi_0) \cdot \bm{\hat{\nu}}(\phi_0) &=& \alpha \iint\displaylimits_{\partial V(\phi_0,\phi_0+\delta\phi)} dS( \bm{\hat{N}} \cdot \bm{\hat{n}}) \>  (\bm{\hat{\nu}}   \cdot  \bm{\hat{n}} ) \nonumber \\
 &+ & \alpha \> \delta\phi \>\> \iint\displaylimits_{X(\phi_0)} dS( \bm{\hat{\nu}} \times \bm{\hat{z}} \cdot \bm{\hat{n}}) \>  (\bm{\hat{\nu}}   \cdot  \bm{\hat{n}} ) \nonumber \\
 \label{eqn: integrated force}
\end{eqnarray}

\noindent where $\bm{\hat{z}}$ is the axis of torsional symmetry (see Fig.  \ref{fig: pillbox}(a)), so that in cylindrical coordinates centered on the axis of symmetry, $\bm{\hat{\nu}} \times \bm{\hat{z}}$ is a unit vector in the radial direction. A detailed derivation of Eq. (\ref{eqn: integrated force}) in the case of general curvilinear coordinates under suitable assumptions is provided in the Supplemental Material \cite{SM1} section IV. Here, we note that in the case of a sample with \emph{high slenderness} (for which the radius of curvature along $\bm{\hat{\nu}}$ is much greater than in the directions perpendicular to it), the second term may be dropped relative to the first term. Once this simplification is made, condition (\ref{eqn: integrated force}) becomes
\begin{equation}
0 \ne \bm{F}(\phi_0) \cdot \bm{\hat{\nu}}(\phi_0) = \alpha \iint\displaylimits_{\partial V(\phi_0,\phi_0+\delta\phi)} dS( \bm{\hat{N}} \cdot \bm{\hat{n}}) \>  (\bm{\hat{\nu}}   \cdot  \bm{\hat{n}} )
\label{eqn: integrated force 2}
\end{equation}
\noindent which we see is met as long as the nematic director $\bf{\hat{n}}$ is not perpendicular to $\bf{\hat{N}}$ or $\bm{\hat{\nu}}$ on all the surfaces bounding the volume element.

\begin{figure}
  \centering
  {\includegraphics[width=.38\textwidth]{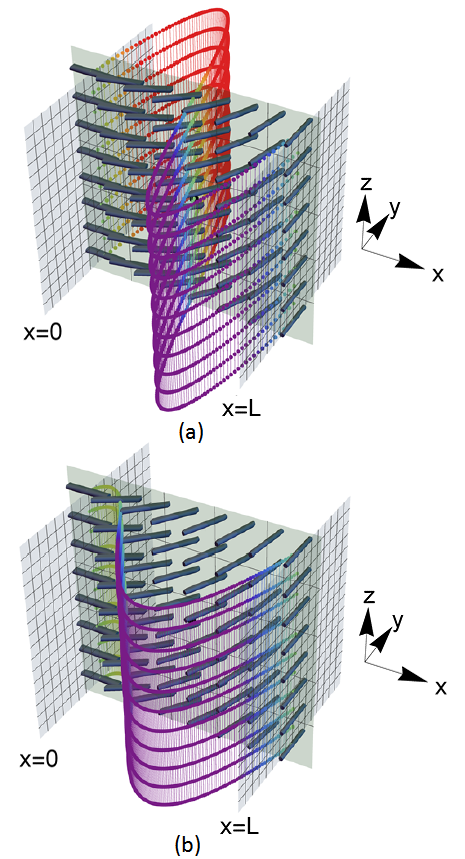}}
  \caption{Color) Flow profile and director field ground state generated with mixed boundary conditions in 2 dimensions (a) in the isotropic case $K_1=K_3$ and (b) in the anisotropic case $K_1 \gg K_3$. Red denotes maximum flow in the $\hat{\bm{y}}$-direction, violet maximum flow in the $-\hat{\bm{y}}$-direction and green no flow. Movies are available in the Supplemental Material \cite{SM}.}
  \label{fig: 2D case}
\end{figure}

To illustrate this criterion with an example, consider an active nematic confined between two infinite parallel plates, one with perpendicular and the other with planar anchoring, shown in Fig.  \ref{fig: 2D case}. The director field and flow profile for this system were determined numerically in Ref. \cite{Marenduzzo2007} and are calculated approximately in Appendix A. Here, we deduce the main features of the flow using simple geometric arguments without carrying out explicit calculations. Firstly, notice that because of the symmetry in the $y-$direction, this system is the high slenderness limit of a similar torsionally symmetric system. This can be seen by giving the system torsional symmetry by revolving the figure about, say, the point $(-R,0)$ in the $x,y-$plane to create an annulus. The high slenderness limit is obtained by sending $R \to \infty$ and recovering Fig. \ref{fig: 2D case}, in which case Eq. (\ref{eqn: integrated force 2}) is exact. However, $\bm{F} \cdot \bm{\hat{\nu}}$ in this cell because $(\bm{\hat{N}} \cdot \bm{\hat{n}})=0$ on one plate and $(\bm{\hat{\nu}}   \cdot  \bm{\hat{n}} )=0$ on the other. Nonetheless, active nematics flow at arbitrary small $\alpha$ in such a mixed alignment cell. This can be explained by applying Eq. (\ref{eqn: integrated force 2}) to either of the two portions of the cell, on opposite sides of the plane (parallel to both walls), whose surface normal $\bf{\hat{N}}$ makes an angle of $\pi/4$ with $\bm{\hat{n}}$. The boundary conditions on $\theta$, and continuity ensure that such a plane exists, though it will not,  for arbitrary and unequal values of the Frank constants $K_{1,2,3}$, be the midplane. According to Eq. (\ref{eqn: integrated force 2}), the resulting active forces in each of the two portions will be non-zero but of opposite sign; hence, the two sides must flow in opposite directions. In the special case of equal Frank constants $K_1=K_2=K_3$, the midplane is the plane on which the surface normal $\bm{\hat{N}}$ makes an angle of $\pi/4$ with $\bm{\hat{n}}$, and the flow in the two halves cancels out, leading to zero net flow in the whole cell. In the generic case of unequal Frank constants, this cancellation does not occur, leading to non-zero net flow, as discussed in Appendix A.

\section{Flow in microchannels with prescribed anchoring angle}

\begin{figure}[b]
  \centering
\includegraphics[width=.48\textwidth]{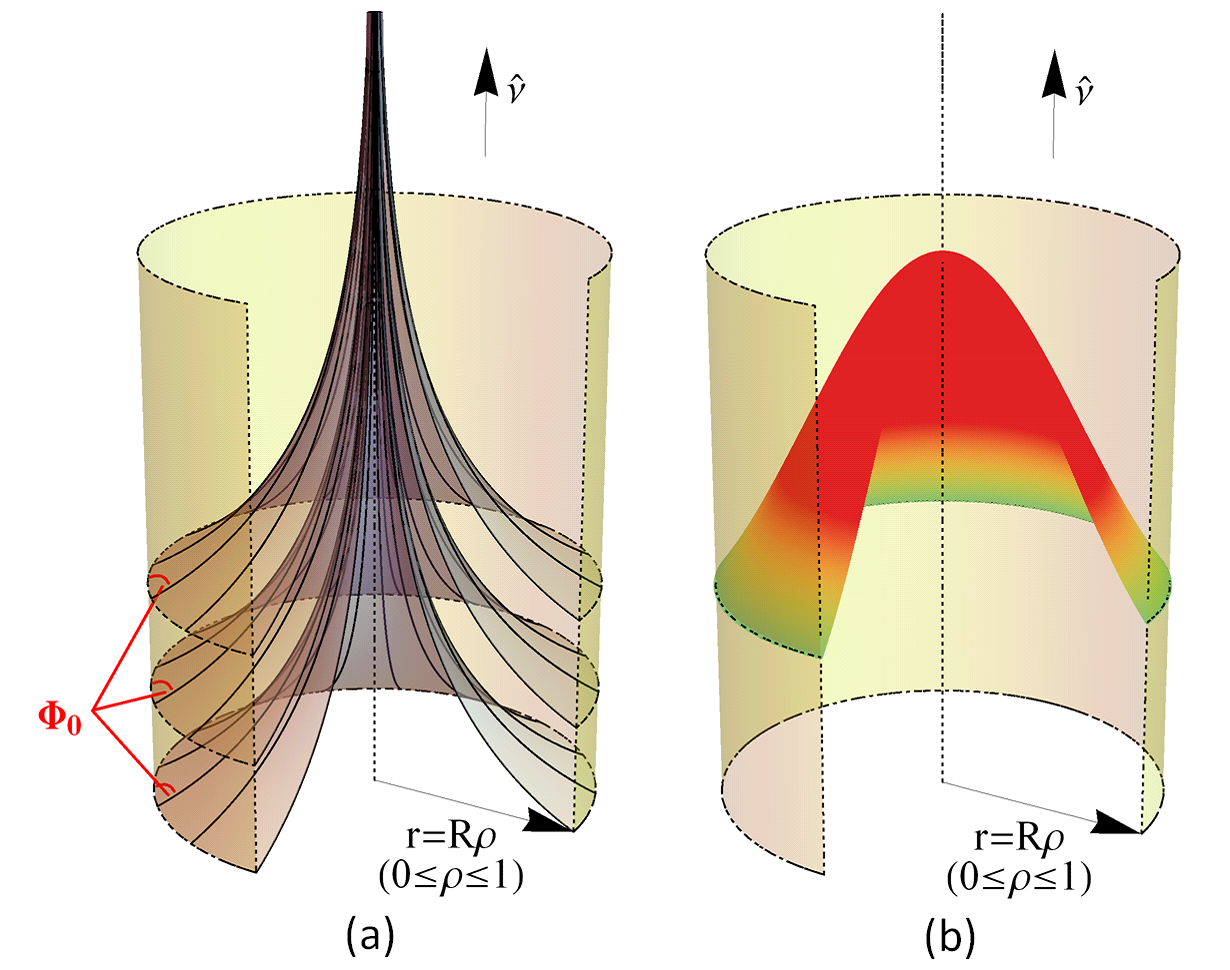}
  \caption{Color) (a) Cylinder director field with constant anchoring angle $\Phi_0$ at the boundary; (b) associated magnitude of the velocity in the $\bm{\hat{\nu}}-$direction in the case of contractile activity ($\alpha<0$).}
  \label{fig: cylinder}
\end{figure}

We now illustrate this criterion for thresholdless active flow in 3D with the simple case of an infinite cylindrical channel with a nematic director field anchored on its boundary at a fixed angle $\Phi_0$ to the axis of the cylinder, as shown in Fig.  \ref{fig: cylinder}. There is full symmetry in the $\theta$-direction as well as the $\bm{\hat{\nu}}-$direction (normally one would call this the $z-$direction, but we have renamed it to be consistent with our notation in section II). Furthermore, since the cylinder can be thought of as an infinitely slender torus, again Eq. (\ref{eqn: integrated force 2}) is exact. 

Even if $\Phi_0=\pi/2$ (homeotropic anchoring), the nematic director configuration that minimizes the Frank free energy gradually ``escapes into the third dimension" \cite{MW, MW'}, becoming aligned with the $\bm{\hat{\nu}}$-axis at the center of the cylinder, as shown in Fig.  \ref{fig: cylinder}. Consider now a different volume, enclosed on the outside by the outer boundary in Fig.  \ref{fig: cylinder}, and on the inside by a concentric inner cylinder, so that $\bm{\hat{N}}$ and $\bm{\hat{\nu}}$ are aligned in the radial and axial direction respectively. Since $\bm{\hat{N}} \cdot \bm{\hat{n}}$ is non-zero on the inner surface, there is a net active force along $\bm{\hat{\nu}}$ which cannot be balanced by pressure gradients, provided that symmetry considerations guarantee that $\bm{\hat{\nu}}\cdot \nabla P = 0$. The same argument can be repeated for any two concentric cylinders inside the channel. We can thus conclude that spontaneous flow along $\bm{\hat{\nu}}$ must occur for arbitrarily small activity, as shown in the right hand panel of Fig.  \ref{fig: cylinder}.

We now proceed to compare the conclusion of the previous argument with an explicit solution of the approximate equations of motion in the frozen director limit. The analytic form for the ground state director field in the one Frank constant approximation (see the left hand panel of Fig. \ref{fig: cylinder}) is given by $\bm{\hat{n}} =  \bm{\hat{\nu}} \cos \Phi - \bm{\hat{\rho}} \sin \Phi$, where $ \rho \equiv r/R$ denotes the dimensionless radial coordinate and $\Phi(\rho)$ satisfies
$\tan \frac{1}{2} \Phi(\rho) = \rho \tan \frac{1}{2} \Phi_0$
\cite{Crawford1991}. The corresponding active force reads

\begin{equation}
\bm{f}_a = \frac{4\alpha \gamma}{R(1+\gamma^2 \rho^2)^3}\left[\bm{\hat{\rho}}(3-\gamma^2 \rho^2) \gamma \rho - \bm{\hat{\nu}}(1-3\gamma^2\rho^2) \right],
\end{equation}

\noindent where $\gamma\equiv \tan{\frac{\Phi_0}{2}}$. In order to solve for the flow $\bm{v}$ in $\nabla P - \eta \nabla^2 \bm{v} = \bm{f}_a$, we use the fact that in a simply connected domain every vector field $\bm{f}_a$ has a unique (up to additive constants) Helmholtz decomposition $\bm{f}_a = \nabla \chi + \nabla \times \bm{A}$ with $\nabla \cdot \bm{A}=0$, provided that on the boundary the normal component of $\nabla \times \bm{A}$ vanishes. Matching the terms respectively with $\nabla P$ and $\nabla^2 \bm{v}$ gives 
\begin{eqnarray}
\nabla P &=& \frac{4\alpha \gamma^2 \rho (3-\gamma^2 \rho^2)}{R(1+\gamma^2 \rho^2)^3} \bm{\hat{\rho}} \nonumber \\
\bm{\Omega}(\rho) &=& \frac{-2\alpha}{\eta}\frac{\gamma\rho(1-\gamma^2\rho^2)}{(1+\gamma^2 \rho^2)^2} \bm{\hat{\theta}}  \nonumber
\end{eqnarray}
\noindent where the vorticity $\bm{\Omega}(\rho)\equiv \nabla \times \bm{v}$, and we have used the identity $\nabla^2 \bm{v} = -\nabla \times \nabla \times \bm{v}$ (which holds since $\nabla \cdot \bm{v}=0$). The relation
\begin{equation}
\iint_{\partial V} dS \> \bm{\hat{\tau}} \cdot \bm{\Omega} \approx -\frac{\alpha}{\eta} \int_{\partial V} dS \> ( \bm{\hat{N}} \cdot \bm{\hat{n_0}}) \>  (\bm{\hat{\nu}}   \cdot  \bm{\hat{n_0}} ),
\end{equation}
\noindent derived in the Supplemental Material \cite{SM1} section V, fixes the constant of integration. Integrating again, and taking into account the no-slip boundary condition for $\bm{v}$, then yields the solution
\begin{equation}
\bm{v} = \frac{-2R\alpha}{\gamma \eta}\left[\frac{1}{1+\gamma^2\rho^2}-\frac{1}{1+\gamma^2}+\frac{1}{2} \log \frac{1+\gamma^2\rho^2}{1+\gamma^2} \right] \bm{\hat{\nu}} \,,
\end{equation}

\noindent which is depicted in Fig.  \ref{fig: cylinder}. The direction of flow is in the positive or negative $\bm{\hat{\nu}}-$direction depending on whether the active forces are extensile ($\alpha<0$) or contractile ($\alpha>0$). Note that the active force changes sign in the bulk if $\Phi_0>\pi/3$, but this is not sufficient to reverse the flow, as can be seen in Fig.  \ref{fig: cylinder}(b).

\section{\label{Sec: Free} Active pumps in a Frederiks cell}

\begin{figure*}
  \centering
\includegraphics[width=0.9\textwidth]{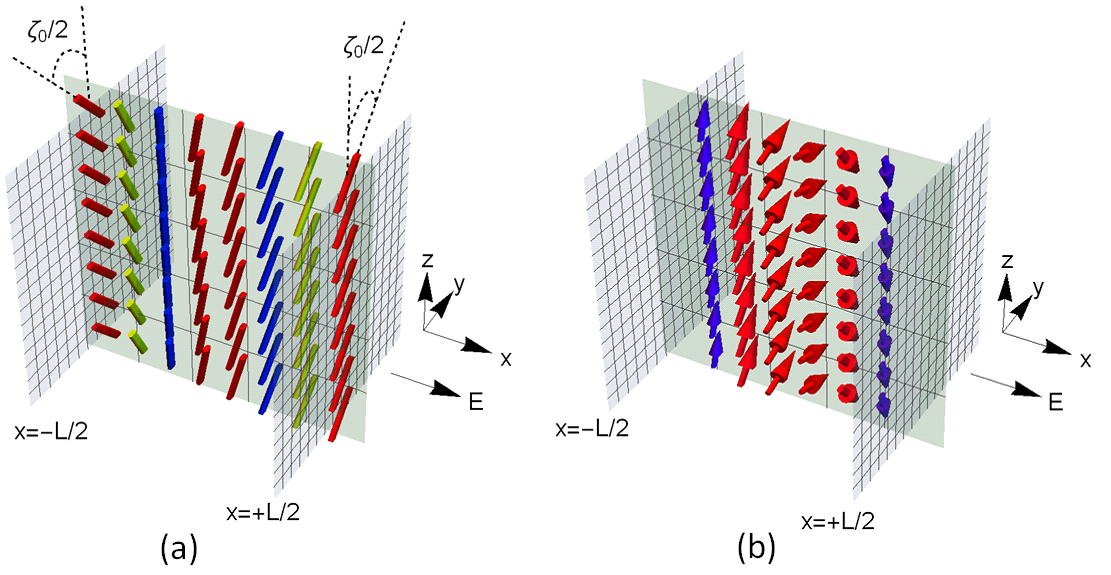}
  \caption{Color) (a) Director field ground state and (b) flow profile for the Frederiks cell. There is net mass transport only in the $y$-direction. A movie is available in the Supplemental Material \cite{SM}.}
  {\label{fig: Frederiks Flow}}
\end{figure*}

We now present the design of an active pump without moving parts based on a nematic Frederiks twist cell. The pump generates a persistent active flow that can be switched on by means of an applied electric field.
As shown in Fig.  \ref{fig: Frederiks Flow}(a), the set-up for the cell is two parallel plates of infinite extent in the $(y,z)$-plane at $x=\pm L/2$.

The plates are prepared with planar (parallel) anchoring but twisted relative to each other by an angle $\zeta_0$. This pure twist nematic distortion leads to a vanishing active force, as noted earlier. However, if a sufficiently large electric field $\bm{E}$ is applied along the $x$-direction, the familiar ``Frederiks instability" \cite{deGennes} can be induced, in which the nematic director tilts towards the $x$-direction inside the cell. This  triggers a spontaneous transverse flow in the $(y,z)$-plane, as can be deduced upon applying Eq. (\ref{eqn: integrated force 2}) (which is again exact using similar arguments to those used for the geometry of Fig.  \ref{fig: 2D case}) to the volume enclosed by two planar boundaries parallel to the plates anywhere inside the cell. If the director is tilted on at least one of the two planar boundaries, then the right hand side of Eq. (\ref{eqn: integrated force 2}) is different from zero, and there is an active force along a direction of symmetry (i.e., the $y$ direction) that cannot be balanced by pressure gradients (in fact, we'll see later that $\nabla \cdot \bm{f}_a=0$ and so $\nabla P=\bm{0}$ everywhere), resulting in flow.

In order to calculate the director field analytically, we parametrize it with the angles $\theta (x)$ and $\zeta (x)$, representing rotation about the $y$- and $x$-axes respectively, so that $\bm{\hat{n}} = \sin \theta (x) \> \bm{\hat{x}} + \cos \theta (x) \> \left[ \sin \zeta (x) \> \bm{\hat{y}} + \cos \zeta (x) \> \bm{\hat{z}} \right]$, with $\theta(\pm L/2) =0$ and $\zeta(\pm L/2)=\pm \zeta_0 /2$. To provide a simplified illustration of the pump design, we assume a single Frank constant $K$. The resulting Euler-Lagrange equations read $K \nabla^2 \bm{\hat{n}} + g n_x \bm{\hat{x}} = \mu (\bm{x}) \bm{\hat{n}}$, where $\mu (\bm{x})$ is the Lagrange multiplier ensuring $\bm{n}^2 =1$ and $g \equiv \epsilon_0 \Delta \chi \> E^2$, where $\Delta \chi$ is the anisotropy in the electric susceptibility and $\epsilon_0$ the permittivity of free space.

As detailed in the Supplemental Material \cite{SM1} section VI, the critical field (above which flow occurs) is given by $g_c = K (\pi/L)^2 (1-(\zeta_0/\pi)^2)$, and writing $g=g_c + \Delta g$, the maximum tilt amplitude $\theta_0$ is related to the incremental field $\Delta g$ close to the transition by $\theta_0^2 = \frac{2 \Delta g}{K k^2 \left(1-\gamma^2 \right)}$, where we've defined $k \equiv\pi/L$ and  $\gamma \equiv \zeta_0/\pi$. Working to $\mathcal{O}(\theta_0)$, the solution for the director field (illustrated in Fig.  \ref{fig: Frederiks Flow}(a)) is $\bm{\hat{n}} = \theta_0 \cos kx \> \bm{\hat{x}} + \bm{\hat{n_0}}$ where $\bm{\hat{n}}_0 \equiv \sin (\gamma kx) \> \bm{\hat{y}} + \cos (\gamma kx) \> \bm{\hat{z}}$. The corresponding active force reads
\begin{eqnarray}
\bm{f}_a &=& -\alpha \theta_0 k \left[ \sin kx \> \bm{\hat{n}}_0 + \gamma \cos kx \> \bm{\hat{x}} \times \bm{\hat{n}}_0 \right],
\end{eqnarray}

\noindent leading to the flow (shown in  Fig.  \ref{fig: Frederiks Flow}(b)):
\begin{eqnarray}
v_y &=& \frac{\alpha \theta_0}{\eta k (1-\gamma^2 )} \Big[\sin \frac{\gamma \pi}{2} - \sin \gamma kx \sin kx \nonumber \\
&& \>\>\>\>\>\>\>\>\>\>\>\>\>\>\>\>\>\>\>\>\>\>\>\>\>\>\>\>\>\>\>\>\>\>\>\>\>\>\>\>\>\>\>\>\>\>\>\> - \gamma \cos \gamma kx \cos kx \Big]\\
v_z &=& \frac{\alpha \theta_0}{\eta k (1-\gamma^2 )} \Big[\frac{2}{\pi}kx \cos \frac{\gamma \pi}{2} - \cos \gamma kx \sin kx \nonumber \\
&& \>\>\>\>\>\>\>\>\>\>\>\>\>\>\>\>\>\>\>\>\>\>\>\>\>\>\>\>\>\>\>\>\>\>\>\>\>\>\>\>\>\>\>\>\>\>\>\> + \gamma \sin \gamma kx \cos kx \Big], \nonumber
\end{eqnarray}

\noindent and $v_x =0$. As previously noted, the flow occurs in the plane transverse to the electric field. Since either $\pm \theta_0$ may be selected, there is spontaneous symmetry breaking when the system selects the sign of $\theta_0$ for the director field, which in turn determines the direction of the flow together with the sign of $\alpha$. There is no net flow in the $z$-direction as $v_z$ is anti-symmetric about $x=0$. However, there is net flow in the $y$-direction, with the maximum attained at $x=0$. 

\vspace{1 pc}

\section{\uppercase{Conclusion}}

In this article, we have addressed what is perhaps the most well known manifestation of active fluids: the onset of spontaneous active flow. Unlike polar active fluids composed of self-propelled particles, active nematics in a uniform configuration do not display net flow unless a critical threshold 
of activity (which is system size dependent) is exceeded. The mechanism of this instability has been extensively studied \cite{actin, actnem}. When activity overcomes the elastic energy it causes a continuous distortion of the alignment of the nematic molecules which in turn generates an active force and hence flow. The typical experimental manifestation of this instability is a chaotic flow marked by the creation and annihilation of defect pairs in the nematic director. 

In this work, we have studied a different mechanism for spontaneous active flow in which flow is induced by special sets of nematic distortions which are {\it imposed} by means of curved substrates, boundaries or external fields, without relying on activity itself to trigger such distortions. In this case, the only role of the active force is to balance frictional or drag forces at steady state (not to compete against elastic forces). As a result, the ensuing flow is laminar, and occurs even for infinitesimally small values of the activity coefficient. Besides highlighting the relation between thresholdless active flow and symmetry breaking of the ground state nematic director, our work may prove of practical interest in the developing field of active nematic microfluidics or experimental studies of ``living liquid crystals" \cite{living} under confinement.

Moreover, the ability to produce such controlled laminar active flows in active nematics makes it possible to study the mechanical response and excitations of materials that break time reversal symmetry by organizing themselves in controlled non-equilibrium steady states. A recent example is the study of topological sound modes propagating within spontaneously flowing polar active liquids under confinement \cite{Souslov2016}, which could now be generalized to active nematics. 
These topological sound modes can be viewed as electrons moving in a pseudo-magnetic field induced by the background flow playing the role of a vector potential. Therefore, our strategies to induce active {\it laminar} flow (well below the instability threshold towards chaotic flow) represent a first step towards engineering well-controlled persistent synthetic gauge fields in active nematics. In future work, we will clarify the topologically protected nature of the active nematic excitations that arise in the presence of such thresholdless laminar flows.

\begin{acknowledgments}
We would like to thank Luca Giomi, Vinzenz Koning, Zvonomir Dogic, Alberto Fernandez Nieves, Jean-Francois Joanny and Oleg Lavrentovich for helpful discussions. RG and VV acknowledge financial support from NWO via a VIDI grant. JT thanks the Max Planck Institute for the Physics of Complex Systems (MPI-PKS), Dresden, Germany;  the Kavli Institute for Theoretical Physics, Santa Barbara, CA; the DITP (NWO zwaartekracht) for financing his stay at the Instituut-Lorentz, Universiteit Leiden and the Department of Bioengineering, Imperial College, London, UK for their hospitality while this work was underway.  He also thanks the  US NSF for support by
awards \#EF-1137815 and \#1006171; and the Simons Foundation for support by award \#225579.
\end{acknowledgments}

\appendix
\renewcommand{\appendixname}{APPENDIX}

\section{\uppercase{Parallel Plates with Mixed Boundary Conditions}}
We consider a two-dimensional system confined between two infinite plates, which has been previously studied numerically \cite{Marenduzzo2007}. The configuration is shown in Fig.  \ref{fig: 2D case}(a) of the main text; the plates are orthogonal to the $x$-axis,  separated by a distance $L$, and prepared with homeotropic alignment (in the $x$-direction) on one plate and planar alignment (in the $y$-direction) on the other. 

We'll begin by demonstrating that for {\it any} values of the Frank constants, this geometry leads to thresholdless flow. Seeking a solution of the form
\begin{equation}
\hat{\bm{n}} = \hat{\bm{x}} \cos \theta(x) + \hat{\bm{y}} \sin \theta(x) ,
\label{bisexgen}
\end{equation} 
to the Euler-Lagrange equations for (\ref{Frank FE}):
\begin{eqnarray}
\mu(\bm{r}) \bm{\hat{n}} = \bm{h}&\equiv&\frac{\delta F}{\delta \nh}= 2(K_{2}-K_{3})\left(\hat{\bm{n}} \cdot \left( \nabla \times \hat{\bm{n}} \right)\right)  \nabla \times \hat{\bm{n}} \nonumber\\ &-&K_{3}\nabla^2\hat{\bm{n}}
+ (K_{3}-K_{2}) \hat{\bm{n}} \times \nabla \left(\hat{\bm{n}}\cdot \nabla \times \hat{\bm{n}} \right)\nonumber\\ &+&(K_{3}-K_{1})\nabla\left(\nabla\cdot\hat{\bm{n}}\right),
\label{hexp}
\end{eqnarray}
where $\mu(\bm{r})$ is a Lagrange multiplier, leads to $ \nabla \times \bm{n}=\hat{\bm{z}} \theta' \cos(\theta)$. Taking this together with (\ref{bisexgen}) implies that the twist vanishes ($\bm{n} \cdot \nabla \times \bm{n}=0$). Using this and (\ref{bisexgen}) in the Euler-Lagrange Eq. (\ref{hexp}) leads to 
\begin{eqnarray}
K_1(\theta'^2\cos\theta+\theta''\sin\theta)\hat{\bm{x}}+ K_3(\theta'^2\sin\theta-\theta''\cos\theta)\hat{\bm{y}} &=& \nonumber\\
\mu(\bm{r})(\hat{\bm{x}} \cos \theta(x) + \hat{\bm{y}} \sin \theta(x)) . \nonumber \\
\label{bisexELG}
\end{eqnarray}
From the $y$ component of Eq. \ref{bisexELG} it follows that 
\begin{eqnarray}
\mu(\bm{r}) \sin \theta = K_3(\theta'^2\sin\theta-\theta''\cos\theta).
\label{bisexmu}
\end{eqnarray}
Using this relation in the $x$ component of (\ref{bisexELG}) gives
\begin{eqnarray}
K_1(\theta'^2\cos\theta+\theta''\sin\theta)\tan \theta = K_3(\theta'^2\sin\theta-\theta''\cos\theta)\,,\nonumber \\
\label{bisexELG2}
\end{eqnarray}
which can be solved for $\theta''$:
\begin{eqnarray}
\theta''={(K_3-K_1)\theta'^2 \sin(2\theta)\over2(K_1\sin^2\theta+K_3\cos^2\theta)}\,.
\label{bisext''}
\end{eqnarray}
From this solution, it is straightforward to show that 
\begin{eqnarray}
\theta'={k\over\sqrt{1+\left({K_3-K_1\over K_3+K_1}\right)\cos(2\theta)}}\,,
\label{bisext'}
\end{eqnarray}
where the constant of integration $k\ne 0$, since the boundary conditions do not allow $\theta(x)$ to be a constant.  Hence, (\ref{bisext'}) implies
that $\theta'\ne 0$ throughout the sample.
Using the original ansatz (\ref{bisexgen}) to calculate the curl of the active force gives, after some algebra,
\begin{eqnarray}
 \nabla \times\bm{f}_a = \alpha \hat{\bm{z}}[\theta'' \cos (2\theta) -2\theta'^2\sin(2\theta)]  \,.
\label{eqn: bisexfc}
\end{eqnarray}
Using our solution (\ref{bisext''}) for $\theta''$ in (\ref{eqn: bisexfc}) gives
\begin{eqnarray}
 \nabla \times\bm{f}_a = \alpha \theta'^2\sin(2\theta)\hat{\bm{z}}\left[\frac{(K_3-K_1) \cos (2\theta) }{2(K_1\sin^2\theta+K_3\cos^2\theta)}-2\right]\,. \nonumber \\
\label{eqn: bisexfc2}
\end{eqnarray}
Since, as we showed earlier, $\theta'\ne 0$ throughout the sample, and the expression in the square brackets is strictly negative, (\ref{eqn: bisexfc2}) implies that $ \nabla \times\bm{f}_a\ne\bm{0}$ throughout the sample, except at the points where $\theta$ is an integer multiple of $\pi/2$. Since $ \nabla \times\bm{f}_a\ne\bm{0}$, there must be flow, as noted in the main text.

From (\ref{bisext'}), we see that the case $K_1=K_3$ (as in the commonly made one Frank constant approximation previously studied numerically \cite{Marenduzzo2007}) is particularly simple, since $\theta'=k$, which implies $\theta=kx+C$ where $C$ is another constant of integration. The constants $C$ and $k$ can be easily determined from the boundary conditions $\theta(x=0)=0$, $\theta(x=L)={\pi\over 2}$, which imply $k=\pi / (2L)$ and $C=0$. Thus
the director field ground state has the solution
\begin{equation}\hat{\bm{n}} = \hat{\bm{x}} \cos kx + \hat{\bm{y}} \sin kx,
\end{equation}
which leads to the active force density: 
\begin{eqnarray}
\bm{f}_a &=& \alpha \left( \bm{n} \nabla \cdot \bm{n} - \bm{n} \times \nabla \times \bm{n} \right) \nonumber \\
&=& \alpha k \left( - \hat{\bm{x}} \sin 2kx + \hat{\bm{y}} \cos 2kx \right).
\label{eqn: f mixed}
\end{eqnarray}
 
To actually determine this  flow $\bm{v}(\bm{r})$, we use the unique Helmholtz decomposition into parts with pure gradient and pure curl. Matching these terms respectively with $\nabla P$ and $\nabla^2 \bm{v}$, as in Eq. (\ref{eqn: simplified Navier Stokes}), and taking into account the no-slip boundary conditions, yields 
\begin{eqnarray}
P &=& \frac{\alpha}{2} \> \cos 2kx \nonumber \\
\bm{v} &=& \frac{\alpha L}{2 \pi \eta} \left( \cos \frac{\pi x}{L} + 2 \frac{x}{L} -1 \right) \> \hat{\bm{y}}.
\end{eqnarray}
As can be seen in Fig.  \ref{fig: 2D case}(a), the flow profile is antisymmetric about the midpoint between the plates at $x=L/2$. Thus there is no net mass transport in this simple example. 
Net mass transport is possible, however, if the alignment angle on the boundary is modified; for example, if it were possible for the right hand plate to be prepared so that the director field made an angle of $\pi/4$ with the normal, then just the left hand half of Fig.  \ref{fig: 2D case} would be realized, with flow now only in the positive $y$-direction.

Net mass transport  also occurs if $K_1 \ne K_3$. Consider the extreme case when $K_1 \gg K_3$; 
then (\ref{bisext'})
implies $\theta'={k\over \sqrt{2}\sin\theta}$, which in turn implies 
\begin{equation}
\cos\theta=C-{k\over \sqrt{2}}x \,,
\label{nosplay}
\end{equation}
where $C$ is another constant of integration. The  boundary conditions $\theta(x=0)=0$, $\theta(x=L)={\pi\over 2}$ now imply $k={\sqrt{2}\over L}$ and $C=1$. Using these in (\ref{nosplay}) then gives
\begin{eqnarray}
\hat{\bm{n}} &=&  \hat{\bm{x}} \left( 1-\frac{x}{L} \right) + \hat{\bm{y}} \sqrt{2 \frac{x}{L} - \left(\frac{x}{L}\right)^2} \\
\bm{f}_a &=&  -\frac{2\alpha}{L} \left( 1-\frac{x}{L} \right)\hat{\bm{x}} +  \alpha \partial_x \left[ \left(1-\frac{x}{L} \right) \sqrt{2\frac{x}{L} - \left(\frac{x}{L} \right) ^2 } \right] \hat{\bm{y}} \nonumber
\end{eqnarray}
The solution  for the pressure $P$ and velocity field $\bm{v}$ is now
\begin{eqnarray}
P &=& -\alpha \left[ 2\frac{x}{L} - \left(\frac{x}{L} \right) ^2 \right] \nonumber \\
\bm{v} &=& \hat{\bm{y}} \> \frac{\alpha L}{3 \eta} \left[ \frac{x}{L} - \left(2\frac{x}{L} - \left(\frac{x}{L} \right) ^2 \right)^{3/2} \right] 
\label{vy}
\end{eqnarray}
\noindent for which there is a net mass transport $J$ in the $y$-direction per unit length in the $z$-direction given by
\begin{equation}
J=\left(\frac{8-3\pi}{48}\right)  \frac{\rho_0 \alpha L^2}{\eta}.
\end{equation}
The velocity field (\ref{vy}) is illustrated in Fig.  \ref{fig: 2D case}(b), which shows that nearly all of the flow is in the negative $y$-direction.

\begin{figure*}[t!!]
  \centering
  {\includegraphics[width=.9\textwidth]{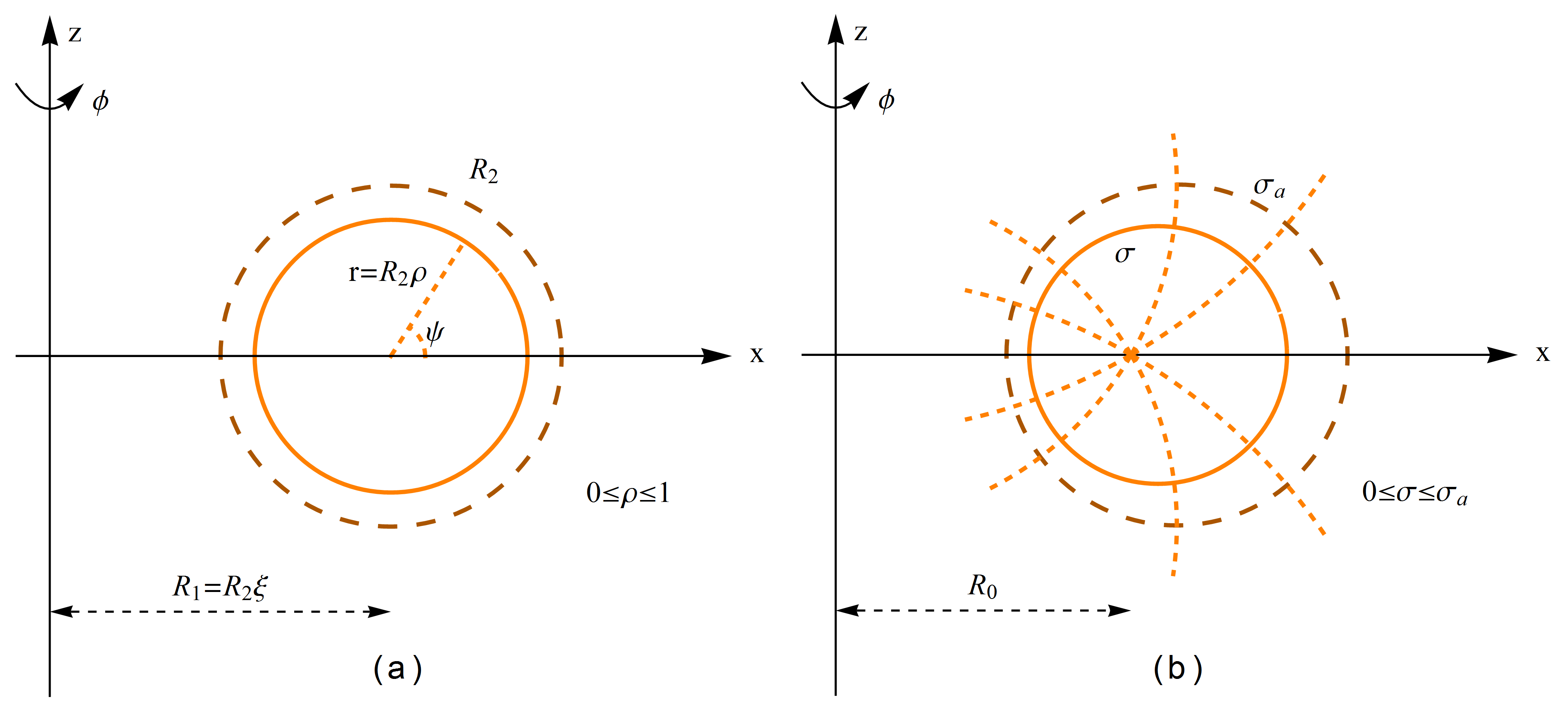}}
  \caption{Color) \emph{Two coordinate systems used to parametrise a toroid:} the plane $y=0$ is shown for the $x\ge 0$ half of the toroid using (a) $(\rho,\psi,\phi)$ coordinates where $\psi$ is the poloidal angle and $\phi$ the azimuthal angle centered at the point $x=R_1, z=0$; and (b) $(\sigma,\tau,\phi)$ coordinates in which $\sigma \in [0,\sigma_a]$ with $\sigma_a<1$. Curves of constant $\tau$ are shown, increasing from $0$ to $2\pi$ in the clockwise direction and converging at the point $x=R_0, z=0$.}
  \label{fig: Toroid_coordinates}
\end{figure*}

\section{\uppercase{Analysis of 3D bulk toroidal nematic}}

We now consider a 3D bulk version of the shell problem dealt with in section IIIB: a bulk toroid with planar anchoring and no-slip on the surface.

\begin{figure}
 \centering
\includegraphics[width=.48\textwidth]{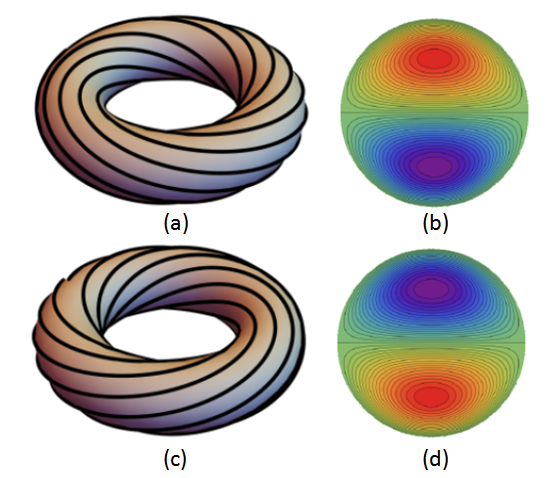}
 \caption{Color) (a) Left chiral ground state ($\omega<0$) of an active nematic liquid and associated flow in cross-section (b). (c) Right chiral ground state ($\omega>0$) of an active nematic liquid and associated flow in cross-section (d). Activity is contractile ($\alpha>0$); red denotes flow in the positive $\phi$-direction, violet denotes flow in the negative $\phi$-direction and green no flow. See Supplemental Material \cite{SM} for a movie of this flow.}
\label{fig: Toroid combined}
\end{figure}

For very slender tori (which are nearly cylinders) the nematic director will be everywhere oriented along $\bm{\hat{\nu}}=\bm{\hat{\phi}}$, the direction of torsional symmetry. Recent experimental and theoretical studies \cite{Pairam2013, Koning2014} have shown that as the aspect ratio of the tori is lowered (i.e., as we move towards ``fatter" tori),  a structural transition to a chiral configuration takes place in the ground state, leading to the twisted nematic texture shown in Fig.  \ref{fig: Toroid combined}.

The following double-twist ansatz has proved effective in capturing the qualitative features of the chiral symmetry breaking transition in the ground state at zero activity \cite{Pairam2013, Koning2014, Davidson2015}:

\begin{equation}
\bm{{\hat n}}=\frac{\omega \xi \rho}{\xi + \rho \cos \psi} \>\> \hat{\bm{\psi}}+\sqrt{1-\left(\frac{\omega \xi \rho}{\xi + \rho \cos \psi}\right)^2} \>\> \hat{\bm{\phi}},
\label{equation: ansatz 3D}
\end{equation}
\noindent where $\omega$ is a variational parameter describing the degree of twist in the director field. In this expression, $(\rho, \psi, \phi)$ are toroidal coordinates (see Fig.  \ref{fig: Toroid_coordinates}(a)),  where $\rho$ is the dimensionless radial coordinate varying between $0$ and $1$, $\psi$ is the poloidal angle, $\phi$ is the toroidal or azimuthal angle and the ``slenderness" of the torus $\xi \equiv R_1/R_2$ is the aspect ratio of its major ($R_2$) and minor ($R_1$) radii. 

To leading order in $1/\xi$, the ground states shown in Fig.  \ref{fig: Toroid combined}(a) are $\omega = \pm 2 \sqrt{\frac{5}{16 \xi^2}-\frac{K_2 - K_{24}}{K_3}}$, provided that the quantity under the square root is positive (otherwise the ground state is the untwisted state $\omega=0$). To $\mathcal{O}(\omega)$, the active force reads:

\begin{equation}
\bm{f}_a = \frac{-\alpha \left(\hat{\bm{\rho}} \cos \psi -\hat{\bm{\psi}} \sin \psi + \hat{\bm{\phi}} \frac{\omega \xi \rho \sin \psi}{\xi + \rho \cos \psi} \right)}{R_2 (\xi + \rho \cos \psi)}.
\label{equation: fa sm}
\end{equation}
In the Supplemental Material \cite{SM1} section III, we show that in the frozen director regime Eq. (\vEOM-\continuity) reduce to
\begin{eqnarray}
\label{eqn: simplified Navier Stokes}
0 &=& - \nabla P + \eta \nabla^2 \bm{v} + \alpha \left( \bm{{\hat n_0}} \cdot \nabla \bm{{\hat n_0}} + \bm{{\hat n_0}}\nabla \cdot \bm{{\hat n_0}}\right) 
 \end{eqnarray}

\noindent with $\nabla \cdot \bm{v} =0$. Using this, and taking the divergence of Eq. (\ref{eqn: simplified Navier Stokes}) implies that $\nabla^2 P=0$. If the pressure is independent of the azimuthal coordinate $\phi$, to $\mathcal{O}(\omega)$, the solution for the pressure
\begin{equation}
\label{eqn pressure}
P=-\alpha \log (\xi + \rho \cos \psi)
\end{equation}
\noindent cancels the source term's $\rho$- and $\psi$-components. If we now write $\bm{v}=\bm{u}(\rho,\psi) + v_{\phi}(\rho,\psi) \hat{\bm{\phi}}$, where $\bm{u}(\rho,\psi)$ is the projection of $\bm{v}$ on the $(\rho,\psi)$-plane, $\bm{u}$ vanishes on the boundary and so throughout the bulk. Thus (\ref{eqn: simplified Navier Stokes}) reduces to
\begin{equation}
\nabla^2 \left(v_{\phi} \hat{\bm{\phi}} \right) = \frac{\alpha \omega \xi \rho \sin \psi}{\eta R_2 (\xi + \rho \cos \psi)} \hat{\bm{\phi}}.
\label{eqn: reduced}
\end{equation}
\noindent with $v_{\phi}=0$ at $\rho=1$. As
\begin{eqnarray}
\nabla^2 \left( v_{\phi}(r,\psi) \hat{\bm{\phi}} \right) &=& \hat{\bm{\phi}} \left( \nabla^2 - \frac{1}{R_2^2(\xi+\rho\cos\psi)^2} \right) v_{\phi} \nonumber \\
&=& \hat{\bm{\phi}} \frac{1}{\cos \phi} \nabla^2 \left( v_{\phi} \cos \phi \right),
\end{eqnarray}
\noindent it is sufficient to solve
\begin{equation}
\label{eq: scalar poisson original}
\nabla^2 \left( v_{\phi}(r,\psi) \cos \phi \right) = \frac{\alpha \omega \xi \rho \sin \psi \cos \phi}{\eta R_2 (\xi + \rho \cos \psi)^2}.
\end{equation}

\noindent In order to do this, we make use of the Green's function for the scalar Poisson equation with vanishing boundary conditions on the surface of a toroid. This is specified in alternative toroidal coordinates $(\sigma, \tau, \phi)$ which are shown in Fig.  \ref{fig: Toroid_coordinates}(b). Mapping our $(\rho,\phi,\psi)$ coordinates to $(\sigma,\tau,\phi)$, firstly we note that
\begin{eqnarray}
& \bm{x} &= R_2 \left[(\xi+\rho \cos \psi) \cos \phi, (\xi+\rho \cos \psi) \sin \phi, \rho \sin \psi \right] \nonumber \\
& \hat{\bm{\rho}} &= \left[\cos \psi \cos \phi, \cos \psi \sin \phi, \sin \psi \right] \\
& \hat{\bm{\psi}} &= \left[-\sin \psi \cos \phi, -\sin \psi \sin \phi, \cos \psi \right] \nonumber \\
& \hat{\bm{\phi}} &= \left[-\sin \phi, \cos \phi, 0 \right] \nonumber \end{eqnarray}
\noindent where $\left[A_x,A_y,A_z \right] \equiv A_x \hat{\bm{x}}+A_y \hat{\bm{y}}+A_z \hat{\bm{z}}$ is shorthand for Cartesian coordinates. The scaling factors in our original coordinate system are
\begin{equation}
h_{\rho} = R_2, \> h_{\phi} = R_2 (\xi + \rho \cos \psi), \> h_{\psi} = R_2 \rho.
\end{equation}
In the alternative coordinate system,
\begin{eqnarray}
& \bm{x} &= \frac{R_0}{1-\sigma \cos \tau} \left[\sqrt{1-\sigma^2} \cos \phi, \sqrt{1-\sigma^2} \sin \phi, -\sigma \sin \tau \right] \nonumber  \\
&&= R_0 (1+\sigma \cos \tau) \left[\cos \phi, \sin \phi, \frac{-\sigma \sin \tau}{(1+\sigma \cos \tau)} \right] + \mathcal{O}(\sigma^2) \nonumber \\
& \hat{\bm{\sigma}} &= \left[\cos \tau \cos \phi, \cos \tau \sin \phi, -\sin \tau \right] + \mathcal{O}(\sigma) \\
& \hat{\bm{\tau}} &= \left[-\sin \tau \cos \phi, -\sin \tau \sin \phi, -\cos \tau \right] + \mathcal{O}(\sigma) \nonumber \\
& \hat{\bm{\phi}} &= \left[-\sin \phi, \cos \phi, 0 \right] \nonumber
\end{eqnarray}
\noindent The surface $\sigma=$ constant ($0 \le \sigma<1$) describes the surface of a torus with major radius $R_1 = R_0 / \sqrt{1-\sigma^2}$ and minor radius $r = R_0 \sigma / \sqrt{1-\sigma^2}$, as can be seen from the fact that $x,y$ and $z$ above satisfy:
\begin{equation}
\left( \sqrt{x^2 + y^2} - \frac{R_0}{\sqrt{1-\sigma^2}} \right)^2 + z^2 = \frac{R_0^2 \sigma^2}{1-\sigma^2}.
\end{equation}
\noindent The scaling factors in the new coordinate system are:
\begin{eqnarray}
h_{\sigma} &=& \frac{R_0}{\sqrt{1-\sigma^2}(1-\sigma \cos \tau)}, \nonumber \\ 
h_{\tau} &=& \frac{R_0 \sigma}{1-\sigma \cos \tau}, \> h_{\phi} = \frac{R_0 \sqrt{1-\sigma^2}}{1-\sigma \cos \tau}.
\end{eqnarray}
\noindent Translating between the original and alternative toroidal coordinate systems,
\begin{eqnarray}
\label{eq: original to alternative}
R_2 = \frac{R_0 \sigma_a}{\sqrt{1-\sigma_a^2}} \>;\>\> \rho = \sqrt \frac{1-\sigma_a^2}{1-\sigma^2} \frac{\sigma}{\sigma_a} \>;\>\> \xi = \frac{R_1}{R_2} = \frac{1}{\sigma_a} \nonumber\\
R_2 (\xi + \rho \cos \psi) = \frac{R_0 \sqrt{1-\sigma^2}}{1-\sigma \cos \tau} \>;\>\> \frac{\rho \sin \psi}{\xi+\rho \cos \psi} = \frac{-\sigma \sin \tau}{\sqrt{1-\sigma^2}}, \nonumber \\
\end{eqnarray}
\noindent and Eq. (\ref{eq: scalar poisson original}) may be expressed in $(\sigma,\tau,\phi)$ coordinates to $\mathcal{O}(\sigma_a)$ as
\begin{equation}
\nabla^2 \left( v_{\phi}(\sigma,\tau) \cos \phi \right) = \frac{-\alpha \omega \sigma \sin \tau \cos \phi}{\eta R_2} + \mathcal{O} (\sigma_a^2).
\end{equation}

In this coordinate system, we may now use the Green's function for the Laplacian vanishing on the surface of a toroid with $\sigma=\sigma_a$, which is given by \cite{Bates1997}:
\begin{eqnarray}
G(\bm{x},\bm{x'})= && \frac{1}{\pi R_0} \sqrt{1- \sigma \cos \tau} \sqrt{1- \sigma' \cos \tau'} \nonumber \\
&& \cdot \sum_{n=0}^{\infty} \sum_{m=0}^{\infty} (-1)^n \epsilon_n \epsilon_m \frac{\Gamma(m-n+1/2)}{\Gamma(m+n+1/2)} g_{mn} \nonumber \\
&& \cdot \cos m(\tau -\tau') \cos n (\phi - \phi')
\end{eqnarray}
\noindent where $g_{mn} \equiv g_{mn}(\sigma',\sigma, \sigma_a)$ is given by
\begin{equation}
g_{mn} \equiv \frac{T_{mn}(\sigma_{<})}{T_{mn}(\sigma_{a})} \cdot \left[T_{mn}(\sigma_{a}) S_{mn}(\sigma_{>}) - T_{mn}(\sigma_{>}) S_{mn}(\sigma_{a}) \right], \nonumber
\end{equation}
\noindent $\epsilon_n$ is $1$ if $n=0$ and $2$ otherwise, and $\sigma_{>}, \sigma_{<}$ denote the higher/lower of $\sigma$ and $\sigma'$ respectively. $T_{mn}(\sigma), S_{mn}(\sigma)$ are toroidal harmonic functions defined as
\begin{eqnarray}
T_{mn}(\sigma) &\equiv & \sigma^{-1/2} Q^n_{m-1/2} (1/\sigma) \>, \nonumber \\
S_{mn}(\sigma) &\equiv & \sigma^{-1/2} P^n_{m-1/2} (1/\sigma), 
\end{eqnarray}
\noindent where the functions $Q^{\lambda}_{\nu} \left(1/\sigma \right)$ and $P^{\lambda}_{\nu} \left(1/\sigma \right)$ are the associated Legendre functions of order $\lambda$ and degree $\nu$.
\vspace{1 pc}
\noindent Applying the Green's function to the source and using the asymptotic forms \cite{ARotenberg1959} $T_{11}(\sigma) \sim -\frac{3\pi}{8\sqrt{2}}\sigma$ and $S_{11} (\sigma) \sim \frac{\sqrt{2}}{\pi} \sigma^{-1}$ for $\sigma \ll 1$, gives to leading order in $\sigma_a = 1/\xi$:
\begin{eqnarray}
v_{\phi} &=& -\frac{4\alpha \omega R_2 \sigma_a}{3\eta} \sin \tau \frac{1}{\sigma_a^3} \int_0^{\sigma_a} d \sigma' \sigma'^2 g_{11} +\mathcal{O} (\sigma_a^2) \nonumber \\
&=& \frac{\alpha \omega R_2 \sigma_a}{8\eta} \sin \tau \>\> s \left(1-s^2 \right) +\mathcal{O} (\sigma_a^2) \nonumber \\
&=& -\frac{\alpha \omega R_2}{8\eta \xi} \sin \psi \>\> \rho \left( 1-\rho^2 \right) + \mathcal{O} (1/\xi^2), 
\end{eqnarray}
\noindent where $s \equiv \sigma / \sigma_a$ and we have used the results (\ref{eq: original to alternative}) to translate back to $(\rho,\phi,\psi)$ coordinates. 

Thus when there is a chiral twisted ground state, the activity creates a flow one way in the $\hat{\bm{\phi}}$-direction in the top half of the toroid and in the opposite direction in the bottom half. The orientation of the flow is determined by the sign of the activity (contractile or extensile) and the chirality of the ground state, as illustrated in Fig.  \ref{fig: Toroid combined}. 

\begin{thebibliography}{42}%

\makeatletter
\providecommand \@ifxundefined [1]{%
 \@ifx{#1\undefined}
}%
\providecommand \@ifnum [1]{%
 \ifnum #1\expandafter \@firstoftwo
 \else \expandafter \@secondoftwo
 \fi
}%
\providecommand \@ifx [1]{%
 \ifx #1\expandafter \@firstoftwo
 \else \expandafter \@secondoftwo
 \fi
}%
\providecommand \natexlab [1]{#1}%
\providecommand \enquote  [1]{``#1''}%
\providecommand \bibnamefont  [1]{#1}%
\providecommand \bibfnamefont [1]{#1}%
\providecommand \citenamefont [1]{#1}%
\providecommand \href@noop [0]{\@secondoftwo}%
\providecommand \href [0]{\begingroup \@sanitize@url \@href}%
\providecommand \@href[1]{\@@startlink{#1}\@@href}%
\providecommand \@@href[1]{\endgroup#1\@@endlink}%
\providecommand \@sanitize@url [0]{\catcode `\\12\catcode `\$12\catcode
  `\&12\catcode `\#12\catcode `\^12\catcode `\_12\catcode `\%12\relax}%
\providecommand \@@startlink[1]{}%
\providecommand \@@endlink[0]{}%
\providecommand \url  [0]{\begingroup\@sanitize@url \@url }%
\providecommand \@url [1]{\endgroup\@href {#1}{\urlprefix }}%
\providecommand \urlprefix  [0]{URL }%
\providecommand \Eprint [0]{\href }%
\providecommand \doibase [0]{http://dx.doi.org/}%
\providecommand \selectlanguage [0]{\@gobble}%
\providecommand \bibinfo  [0]{\@secondoftwo}%
\providecommand \bibfield  [0]{\@secondoftwo}%
\providecommand \translation [1]{[#1]}%
\providecommand \BibitemOpen [0]{}%
\providecommand \bibitemStop [0]{}%
\providecommand \bibitemNoStop [0]{.\EOS\space}%
\providecommand \EOS [0]{\spacefactor3000\relax}%
\providecommand \BibitemShut  [1]{\csname bibitem#1\endcsname}%
\let\auto@bib@innerbib\@empty

\bibitem [{\citenamefont {Marchetti}\ \emph {et~al.}(2013)\citenamefont
  {Marchetti}, \citenamefont {Joanny}, \citenamefont {Ramaswamy}, \citenamefont
  {Liverpool}, \citenamefont {Prost}, \citenamefont {Rao},\ and\ \citenamefont
  {Simha}}]{Marchetti2013}%
  \BibitemOpen
  \bibfield  {author} {\bibinfo {author} {\bibfnamefont {M.~C.}\ \bibnamefont
  {Marchetti}}, \bibinfo {author} {\bibfnamefont {J.~F.}\ \bibnamefont
  {Joanny}}, \bibinfo {author} {\bibfnamefont {S.}~\bibnamefont {Ramaswamy}},
  \bibinfo {author} {\bibfnamefont {T.~B.}\ \bibnamefont {Liverpool}}, \bibinfo
  {author} {\bibfnamefont {J.}~\bibnamefont {Prost}}, \bibinfo {author}
  {\bibfnamefont {M.}\ \bibnamefont {Rao}}, \ and\ \bibinfo {author}
  {\bibfnamefont {R.~Aditi}\ \bibnamefont {Simha}},\ }\bibfield  {title}
  {\emph {\bibinfo {title} {{Hydrodynamics of soft active matter}},}\ }\href
  {\doibase 10.1103/RevModPhys.85.1143} {\bibfield  {journal} {\bibinfo
  {journal} {Reviews of Modern Physics}\ }\textbf {\bibinfo {volume} {85}},\
  \bibinfo {pages} {1143--1189} (\bibinfo {year} {2013})}\BibitemShut {NoStop}%
\bibitem [{\citenamefont {Voituriez}\ \emph {et~al.}(2005)\citenamefont
  {Voituriez}, \citenamefont {Joanny},\ and\ \citenamefont {Prost}}]{actin}%
  \BibitemOpen
  \bibfield  {author} {\bibinfo {author} {\bibfnamefont {R.}~\bibnamefont
  {Voituriez}}, \bibinfo {author} {\bibfnamefont {J.~F.}\ \bibnamefont
  {Joanny}}, \ and\ \bibinfo {author} {\bibfnamefont {J.}~\bibnamefont
  {Prost}},\ }\bibfield  {title} {\emph {\bibinfo {title} {Spontaneous flow
  transition in active polar gels},}\ }\href@noop {} {\bibfield  {journal}
  {\bibinfo  {journal} {Europhysics Letters}\ }\textbf {\bibinfo {volume}
  {70(3)}},\ \bibinfo {pages} {404} (\bibinfo {year} {2005})}\BibitemShut
  {NoStop}%
\bibitem [{\citenamefont {{Kruse, K.}}\ \emph {et~al.}(2005)\citenamefont
  {{Kruse, K.}}, \citenamefont {{Joanny, J. F.}}, \citenamefont {{Julicher,
  F.}}, \citenamefont {{Prost, J.}},\ and\ \citenamefont {{Sekimoto,
  K.}}}]{microtub}%
  \BibitemOpen
  \bibfield  {author} {\bibinfo {author} {\bibnamefont {{K. Kruse}}}, \bibinfo
  {author} {\bibnamefont {{J.~F. Joanny}}}, \bibinfo {author} {\bibnamefont
  {{F. Jülicher}}}, \bibinfo {author} {\bibnamefont {{J. Prost}}}, \ and\
  \bibinfo {author} {\bibnamefont {{K. Sekimoto}}},\ }\bibfield  {title}
  {\emph {\bibinfo {title} {Generic theory of active polar gels: a paradigm
  for cytoskeletal dynamics},}\ }\href {\doibase 10.1140/epje/e2005-00002-5}
  {\bibfield  {journal} {\bibinfo  {journal} {Eur. Phys. J. E}\ }\textbf
  {\bibinfo {volume} {16}},\ \bibinfo {pages} {5--16} (\bibinfo {year}
  {2005})}\BibitemShut {NoStop}%
\bibitem [{\citenamefont {Paxton}\ \emph {et~al.}(2006)\citenamefont {Paxton},
  \citenamefont {Bake}, \citenamefont {Kline}, \citenamefont {Wang},
  \citenamefont {Mallouk},\ and\ \citenamefont {Sen}}]{Janus}%
  \BibitemOpen
  \bibfield  {author} {\bibinfo {author} {\bibfnamefont {W.~F.}\
  \bibnamefont {Paxton}}, \bibinfo {author} {\bibfnamefont {P.~T.}\
  \bibnamefont {Bake}}, \bibinfo {author} {\bibfnamefont {T.~R.}\
  \bibnamefont {Kline}}, \bibinfo {author} {\bibfnamefont {Y.}\ \bibnamefont
  {Wang}}, \bibinfo {author} {\bibfnamefont {T.~E.}\ \bibnamefont
  {Mallouk}}, \ and\ \bibinfo {author} {\bibfnamefont {A.}\ \bibnamefont
  {Sen}},\ }\bibfield  {title} {\emph {\bibinfo {title} {Catalytically
  induced electrokinetics for motors and micropumps},}\ }\href@noop {}
  {\bibfield  {journal} {\bibinfo  {journal} {Journal of the American Chemical
  Society}\ }\textbf {\bibinfo {volume} {128}},\ \bibinfo {pages}
  {14881--14888} (\bibinfo {year} {2006})}\BibitemShut {NoStop}%
\bibitem [{\citenamefont {Reynolds}(1987)}]{boids1}%
  \BibitemOpen
  \bibfield  {author} {\bibinfo {author} {\bibfnamefont {C.~W.}\
  \bibnamefont {Reynolds}},\ }\bibfield  {title} {\emph {\bibinfo {title}
  {Flocks, herds and schools: A distributed behavioral model},}\ }\href
  {\doibase 10.1145/37402.37406} {\bibfield  {journal} {\bibinfo  {journal}
  {Comput. Graph.}\ }\textbf {\bibinfo {volume} {21(4)}},\ \bibinfo
  {pages} {25--34} (\bibinfo {year} {1987})}\BibitemShut {NoStop}%
\bibitem [{\citenamefont {Deneubourg}\ and\ \citenamefont
  {Goss}(1989)}]{boids2}%
  \BibitemOpen
  \bibfield  {author} {\bibinfo {author} {\bibfnamefont {J.L.}\ \bibnamefont
  {Deneubourg}}\ and\ \bibinfo {author} {\bibfnamefont {S.}~\bibnamefont
  {Goss}},\ }\bibfield  {title} {\emph {\bibinfo {title} {Collective
  patterns and decision-making},}\ }\href@noop {} {\bibfield  {journal}
  {\bibinfo  {journal} {Ethology Ecology \& Evolution}\ }\textbf {\bibinfo
  {volume} {1}},\ \bibinfo {pages} {295--311} (\bibinfo {year}
  {1989})}\BibitemShut {NoStop}%
\bibitem [{\citenamefont {Huth}\ and\ \citenamefont {Wissel}(1990)}]{boids3}%
  \BibitemOpen
  \bibfield  {author} {\bibinfo {author} {\bibfnamefont {A.}~\bibnamefont
  {Huth}}\ and\ \bibinfo {author} {\bibfnamefont {C.}~\bibnamefont {Wissel}},\
  }\bibfield  {title} {\emph {\bibinfo {title} {The movement of fish
  schools: a simulation model},}\ }\href@noop {} {\bibfield  {journal}
  {\bibinfo  {journal} {in {\em Biological Motion}}\ ,\ \bibinfo {pages}
  {577--590}} (\bibinfo {year} {1990})}\BibitemShut {NoStop}%
\bibitem [{\citenamefont {Partridge}(1982)}]{boids4}%
  \BibitemOpen
  \bibfield  {author} {\bibinfo {author} {\bibfnamefont {B.~L.}\ \bibnamefont
  {Partridge}},\ }\bibfield  {title} {\emph {\bibinfo {title} {The structure
  and function of fish schools},}\ }\href@noop {} {\bibfield  {journal}
  {\bibinfo  {journal} {Scientific American}\ ,\ \bibinfo {pages} {114--123}}
  (\bibinfo {year} {1982})}\BibitemShut {NoStop}%
\bibitem [{\citenamefont {Tunstrom}\ \emph {et~al.}(2013)\citenamefont
  {Tunstrom}, \citenamefont {Katz}, \citenamefont {Ioannou}, \citenamefont
  {Huepe}, \citenamefont {Lutz},\ and\ \citenamefont {Couzin}}]{couzin}%
  \BibitemOpen
  \bibfield  {author} {\bibinfo {author} {\bibfnamefont {K.}~\bibnamefont
  {Tunstrom}}, \bibinfo {author} {\bibfnamefont {Y.}~\bibnamefont {Katz}},
  \bibinfo {author} {\bibfnamefont {C.~C.}\ \bibnamefont {Ioannou}}, \bibinfo
  {author} {\bibfnamefont {C.}~\bibnamefont {Huepe}}, \bibinfo {author}
  {\bibfnamefont {M.}~\bibnamefont {Lutz}}, \ and\ \bibinfo {author}
  {\bibfnamefont {I.~D.}\ \bibnamefont {Couzin}},\ }\bibfield  {title}
  {\emph {\bibinfo {title} {Collective states, multistability and
  transitional behavior in schooling fish},}\ }\href@noop {} {\bibfield
  {journal} {\bibinfo  {journal} {PLoS Computational Biology}\ }\textbf
  {\bibinfo {volume} {9(2)}},\ \bibinfo {pages} {1002915} (\bibinfo {year}
  {2013})}\BibitemShut {NoStop}%
\bibitem [{\citenamefont {Loomis}(1982)}]{dictyo1}%
  \BibitemOpen
  \bibfield  {author} {\bibinfo {author} {\bibfnamefont {W.}~\bibnamefont
  {Loomis}},\ }\href@noop {} {\emph {\bibinfo {title} {{The Development of
  Dictyostelium discoideum}}}}\ (\bibinfo  {publisher} {Academic, New York},\
  \bibinfo {year} {1982})\BibitemShut {NoStop}%
\bibitem [{\citenamefont {Bonner}(1967)}]{dictyo2}%
  \BibitemOpen
  \bibfield  {author} {\bibinfo {author} {\bibfnamefont {J.T.}\ \bibnamefont
  {Bonner}},\ }\href@noop {} {\emph {\bibinfo {title} {{The Cellular Slime
  Molds}}}}\ (\bibinfo  {publisher} {Princeton University Press},\ \bibinfo
  {year} {1967})\BibitemShut {NoStop}%
\bibitem [{\citenamefont {Attanasi}\ \emph {et~al.}(2014)\citenamefont
  {Attanasi}, \citenamefont {Cavagna}, \citenamefont {Del~Castello},
  \citenamefont {Giardina}, \citenamefont {Melillo}, \citenamefont {Parisi},
  \citenamefont {Pohl}, \citenamefont {Rossaro}, \citenamefont {Shen},
  \citenamefont {Silvestri},\ and\ \citenamefont {Viale}}]{midges}%
  \BibitemOpen
  \bibfield  {author} {\bibinfo {author} {\bibfnamefont {A.}\
  \bibnamefont {Attanasi}}, \bibinfo {author} {\bibfnamefont {A.}\
  \bibnamefont {Cavagna}}, \bibinfo {author} {\bibfnamefont {L.}\
  \bibnamefont {Del~Castello}}, \bibinfo {author} {\bibfnamefont {I.}\
  \bibnamefont {Giardina}}, \bibinfo {author} {\bibfnamefont {S.}\
  \bibnamefont {Melillo}}, \bibinfo {author} {\bibfnamefont {L.}\
  \bibnamefont {Parisi}}, \bibinfo {author} {\bibfnamefont {O.}\
  \bibnamefont {Pohl}}, \bibinfo {author} {\bibfnamefont {B.}\ \bibnamefont
  {Rossaro}}, \bibinfo {author} {\bibfnamefont {E.}\ \bibnamefont {Shen}},
  \bibinfo {author} {\bibfnamefont {E.}\ \bibnamefont {Silvestri}}, \ and\
  \bibinfo {author} {\bibfnamefont {M.}\ \bibnamefont {Viale}},\
  }\bibfield  {title} {\emph {\bibinfo {title} {Collective behaviour without
  collective order in wild swarms of midges},}\ }\href {\doibase
  10.1371/journal.pcbi.1003697} {\bibfield  {journal} {\bibinfo  {journal}
  {PLoS Comput Biol}\ }\textbf {\bibinfo {volume} {10}},\ \bibinfo {pages}
  {1--10} (\bibinfo {year} {2014})}\BibitemShut {NoStop}%
\bibitem [{\citenamefont {Zhou}\ \emph {et~al.}(2014)\citenamefont {Zhou},
  \citenamefont {Sokolov}, \citenamefont {Lavrentovich},\ and\ \citenamefont
  {Aranson}}]{living}%
  \BibitemOpen
  \bibfield  {author} {\bibinfo {author} {\bibfnamefont {S.}~\bibnamefont
  {Zhou}}, \bibinfo {author} {\bibfnamefont {A.}~\bibnamefont {Sokolov}},
  \bibinfo {author} {\bibfnamefont {O.~D.}\ \bibnamefont {Lavrentovich}}, \
  and\ \bibinfo {author} {\bibfnamefont {I.~S.}\ \bibnamefont {Aranson}},\
  }\bibfield  {title} {\emph {\bibinfo {title} {{Living liquid crystals}},}\
  }\href@noop {} {\bibfield  {journal} {\bibinfo  {journal} {Proceedings of the
  National Academy of Sciences}\ }\textbf {\bibinfo {volume} {111}},\ \bibinfo
  {pages} {1265} (\bibinfo {year} {2014})}\BibitemShut {NoStop}%
\bibitem [{\citenamefont {Toner}\ and\ \citenamefont {h~Tu}(1995)}]{TT1}%
  \BibitemOpen
  \bibfield  {author} {\bibinfo {author} {\bibfnamefont {J.}~\bibnamefont
  {Toner}}\ and\ \bibinfo {author} {\bibfnamefont {Y.}~\bibnamefont {Tu}},\
  }\bibfield  {title} {\emph {\bibinfo {title} {Long-range order in a
  two-dimensional dynamical $xy$ model: How birds fly together},}\ }\href@noop
  {} {\bibfield  {journal} {\bibinfo  {journal} {Phys.\ Rev.\ Lett.}\ }\textbf
  {\bibinfo {volume} {\bf{75}}},\ \bibinfo {pages} {4326} (\bibinfo {year}
  {1995})}\BibitemShut {NoStop}%
\bibitem [{\citenamefont {Tu}\ \emph {et~al.}(1998)\citenamefont {Tu},
  \citenamefont {Toner},\ and\ \citenamefont {Ulm}}]{TT2}%
  \BibitemOpen
  \bibfield  {author} {\bibinfo {author} {\bibfnamefont {Y.}~\bibnamefont
  {Tu}}, \bibinfo {author} {\bibfnamefont {J.}~\bibnamefont {Toner}}, \ and\
  \bibinfo {author} {\bibfnamefont {M.}~\bibnamefont {Ulm}},\ }\bibfield
  {title} {\emph {\bibinfo {title} {Sound waves and the absence of galilean
  invariance in flocks},}\ }\href@noop {} {\bibfield  {journal} {\bibinfo
  {journal} {Phys.\ Rev.\ Lett.}\ }\textbf {\bibinfo {volume} {\bf{80}}},\
  \bibinfo {pages} {4819} (\bibinfo {year} {1998})}\BibitemShut {NoStop}%
\bibitem [{\citenamefont {Toner}\ and\ \citenamefont {Tu}(1998)}]{TT3}%
  \BibitemOpen
  \bibfield  {author} {\bibinfo {author} {\bibfnamefont {J.}~\bibnamefont
  {Toner}}\ and\ \bibinfo {author} {\bibfnamefont {Y.}~\bibnamefont {Tu}},\
  }\bibfield  {title} {\emph {\bibinfo {title} {Flocks, herds, and schools:
  A quantitative theory of flocking},}\ }\href@noop {} {\bibfield  {journal}
  {\bibinfo  {journal} {Phys.\ Rev.\ E}\ }\textbf {\bibinfo {volume}
  {\bf{58}}},\ \bibinfo {pages} {4828} (\bibinfo {year} {1998})}\BibitemShut
  {NoStop}%
\bibitem [{\citenamefont {Toner}\ \emph {et~al.}(2005)\citenamefont {Toner},
  \citenamefont {Tu},\ and\ \citenamefont {Ramaswamy}}]{TT4}%
  \BibitemOpen
  \bibfield  {author} {\bibinfo {author} {\bibfnamefont {J.}~\bibnamefont
  {Toner}}, \bibinfo {author} {\bibfnamefont {Y.}~\bibnamefont {Tu}}, \ and\
  \bibinfo {author} {\bibfnamefont {S.}~\bibnamefont {Ramaswamy}},\ }\bibfield
  {title} {\emph {\bibinfo {title} {Hydrodynamics and phases of flocks},}\
  }\href@noop {} {\bibfield  {journal} {\bibinfo  {journal} {Ann.\ Phys.}\
  }\textbf {\bibinfo {volume} {\bf{318}}},\ \bibinfo {pages} {179} (\bibinfo
  {year} {2005})}\BibitemShut {NoStop}%
\bibitem [{\citenamefont {Simha}\ and\ \citenamefont
  {Ramaswamy}(2002)}]{actnem}%
  \BibitemOpen
  \bibfield  {author} {\bibinfo {author} {\bibfnamefont {R.~A.}\ \bibnamefont
  {Simha}}\ and\ \bibinfo {author} {\bibfnamefont {S.}~\bibnamefont
  {Ramaswamy}},\ }\bibfield  {title} {\emph {\bibinfo {title} {Hydrodynamic fluctuations and instabilities in ordered suspensions of self-propelled particles},}\ }\href@noop {} {\bibfield  {journal} {\bibinfo
  {journal} {Phys.\ Rev.\ Lett.}\ }\textbf {\bibinfo {volume} {\bf{89}}},\
  \bibinfo {pages} {058101} (\bibinfo {year} {2002})}\BibitemShut {NoStop}%
  \bibitem [{\citenamefont {{Souslov}}\ \emph {et~al.}(2016)\citenamefont
  {{Souslov}}, \citenamefont {{van Zuiden}}, \citenamefont {{Bartolo}},\ and\
  \citenamefont {{Vitelli}}}]{Souslov2016}%
  \BibitemOpen
  \bibfield  {author} {\bibinfo {author} {\bibfnamefont {A.}~\bibnamefont
  {{Souslov}}}, \bibinfo {author} {\bibfnamefont {B.~C.}\ \bibnamefont {{van
  Zuiden}}}, \bibinfo {author} {\bibfnamefont {D.}~\bibnamefont {{Bartolo}}}, \
  and\ \bibinfo {author} {\bibfnamefont {V.}~\bibnamefont {{Vitelli}}},\
  }\bibfield  {title} {\emph {\bibinfo {title} {{Topological sound in
  active-liquid metamaterials}},}\ }\href@noop {} {\bibfield  {journal}
  {\bibinfo  {journal} {ArXiv e-prints}\ } (\bibinfo {year} {2016})},\ \Eprint
  {http://arxiv.org/abs/1610.06873} {arXiv:1610.06873 [cond-mat.soft]}
  \BibitemShut {NoStop}%
\bibitem [{\citenamefont {Keber}\ \emph {et~al.}(2014)\citenamefont {Keber},
  \citenamefont {Loiseau}, \citenamefont {Sanchez}, \citenamefont {DeCamp},
  \citenamefont {Giomi}, \citenamefont {Bowick}, \citenamefont {Marchetti},
  \citenamefont {Dogic},\ and\ \citenamefont {Bausch}}]{Keber2014}%
  \BibitemOpen
  \bibfield  {author} {\bibinfo {author} {\bibfnamefont {F.~C.}\ \bibnamefont
  {Keber}}, \bibinfo {author} {\bibfnamefont {E.}\ \bibnamefont
  {Loiseau}}, \bibinfo {author} {\bibfnamefont {T.}\ \bibnamefont {Sanchez}},
  \bibinfo {author} {\bibfnamefont {S.~J.}\ \bibnamefont {DeCamp}},
  \bibinfo {author} {\bibfnamefont {L.}\ \bibnamefont {Giomi}}, \bibinfo
  {author} {\bibfnamefont {M.~J.}\ \bibnamefont {Bowick}}, \bibinfo {author}
  {\bibfnamefont {M.~C.}\ \bibnamefont {Marchetti}}, \bibinfo {author}
  {\bibfnamefont {Z.}\ \bibnamefont {Dogic}}, \ and\ \bibinfo {author}
  {\bibfnamefont {A.~R.}\ \bibnamefont {Bausch}},\ }\bibfield  {title}
  {\emph {\bibinfo {title} {{Topology and dynamics of active nematic
  vesicles}},}\ }\href {\doibase 10.1126/science.1254784} {\bibfield  {journal}
  {\bibinfo  {journal} {Science}\ }\textbf {\bibinfo {volume} {345}},\ \bibinfo
  {pages} {1135--1139} (\bibinfo {year} {2014})}\BibitemShut {NoStop}%
\bibitem [{\citenamefont {Giomi}\ \emph {et~al.}(2013)\citenamefont {Giomi},
  \citenamefont {Bowick}, \citenamefont {Ma},\ and\ \citenamefont
  {Marchetti}}]{Giomi2013}%
  \BibitemOpen
  \bibfield  {author} {\bibinfo {author} {\bibfnamefont {L.}\ \bibnamefont
  {Giomi}}, \bibinfo {author} {\bibfnamefont {M.~J.}\ \bibnamefont {Bowick}},
  \bibinfo {author} {\bibfnamefont {X.}~\bibnamefont {Ma}}, \ and\ \bibinfo
  {author} {\bibfnamefont {M.~C.}\ \bibnamefont {Marchetti}},\ }\bibfield
  {title} {\emph {\bibinfo {title} {{Defect Annihilation and Proliferation
  in Active Nematics}},}\ }\href@noop {} {\bibfield  {journal} {\bibinfo
  {journal} {Phys. Rev. Lett.}\ }\textbf {\bibinfo {volume} {110}},\
  \bibinfo {pages} {228101} (\bibinfo {year} {2013})}\BibitemShut {NoStop}%
\bibitem [{\citenamefont {Giomi}(2015)}]{Giomi2014}%
  \BibitemOpen
  \bibfield  {author} {\bibinfo {author} {\bibfnamefont {L.}\ \bibnamefont
  {Giomi}},\ }\bibfield  {title} {\emph {\bibinfo {title} {Geometry and
  topology of turbulence in active nematics},}\ }\href {\doibase
  10.1103/PhysRevX.5.031003} {\bibfield  {journal} {\bibinfo  {journal} {Phys.
  Rev. X}\ }\textbf {\bibinfo {volume} {5}},\ \bibinfo {pages} {031003}
  (\bibinfo {year} {2015})}\BibitemShut {NoStop}%
\bibitem [{\citenamefont {Sanchez}\ \emph {et~al.}(2012)\citenamefont
  {Sanchez}, \citenamefont {Chen}, \citenamefont {DeCamp}, \citenamefont
  {Heymann},\ and\ \citenamefont {Dogic}}]{Sanchez2012}%
  \BibitemOpen
  \bibfield  {author} {\bibinfo {author} {\bibfnamefont {T.}\ \bibnamefont
  {Sanchez}}, \bibinfo {author} {\bibfnamefont {D.~T.~N.}\ \bibnamefont
  {Chen}}, \bibinfo {author} {\bibfnamefont {S.~J.}\ \bibnamefont
  {DeCamp}}, \bibinfo {author} {\bibfnamefont {M.}\ \bibnamefont
  {Heymann}}, \ and\ \bibinfo {author} {\bibfnamefont {Z.}\ \bibnamefont
  {Dogic}},\ }\bibfield  {title} {\emph {\bibinfo {title} {{Spontaneous
  motion in hierarchically assembled active matter.}}}\ }\href {\doibase
  10.1038/nature11591} {\bibfield  {journal} {\bibinfo  {journal} {Nature}\
  }\textbf {\bibinfo {volume} {491}},\ \bibinfo {pages} {431--4} (\bibinfo
  {year} {2012})}\BibitemShut {NoStop}%
\bibitem [{\citenamefont {Ravnik}\ and\ \citenamefont
  {Yeomans}(2013)}]{Ravnik2013}%
  \BibitemOpen
  \bibfield  {author} {\bibinfo {author} {\bibfnamefont {M.}\ \bibnamefont
  {Ravnik}}\ and\ \bibinfo {author} {\bibfnamefont {J.M.}\ \bibnamefont
  {Yeomans}},\ }\bibfield  {title} {\emph {\bibinfo {title} {{Confined
  Active Nematic Flow in Cylindrical Capillaries}},}\ }\href {\doibase
  10.1103/PhysRevLett.110.026001} {\bibfield  {journal} {\bibinfo  {journal}
  {Phys. Rev. Lett.}\ }\textbf {\bibinfo {volume} {110}},\ \bibinfo
  {pages} {026001} (\bibinfo {year} {2013})}\BibitemShut {NoStop}%
\bibitem [{\citenamefont {Marenduzzo}\ \emph {et~al.}(2007)\citenamefont
  {Marenduzzo}, \citenamefont {Orlandini}, \citenamefont {Cates},\ and\
  \citenamefont {Yeomans}}]{Marenduzzo2007}%
  \BibitemOpen
  \bibfield  {author} {\bibinfo {author} {\bibfnamefont {D.}~\bibnamefont
  {Marenduzzo}}, \bibinfo {author} {\bibfnamefont {E.}~\bibnamefont
  {Orlandini}}, \bibinfo {author} {\bibfnamefont {M.~E.}\ \bibnamefont
  {Cates}}, \ and\ \bibinfo {author} {\bibfnamefont {J.~M.}\ \bibnamefont
  {Yeomans}},\ }\bibfield  {title} {\emph {\bibinfo {title} {Steady-state
  hydrodynamic instabilities of active liquid crystals: Hybrid lattice
  boltzmann simulations},}\ }\href@noop {} {\bibfield  {journal} {\bibinfo
  {journal} {Physics Review E}\ }\textbf {\bibinfo {volume} {\bf{76}}},\
  \bibinfo {pages} {031921} (\bibinfo {year} {2007})}\BibitemShut {NoStop}%
\bibitem [{\citenamefont {Giomi}\ \emph {et~al.}(2008)\citenamefont {Giomi},
  \citenamefont {Marchetti},\ and\ \citenamefont {Liverpool}}]{GiomiPRL2008}%
  \BibitemOpen
  \bibfield  {author} {\bibinfo {author} {\bibfnamefont {L.}\ \bibnamefont
  {Giomi}}, \bibinfo {author} {\bibfnamefont {M.~C.}\ \bibnamefont
  {Marchetti}}, \ and\ \bibinfo {author} {\bibfnamefont {T.~B.}\
  \bibnamefont {Liverpool}},\ }\bibfield  {title} {\emph {\bibinfo {title}
  {Complex spontaneous flows and concentration banding in active polar
  films},}\ }\href {\doibase 10.1103/PhysRevLett.101.198101} {\bibfield
  {journal} {\bibinfo  {journal} {Phys. Rev. Lett.}\ }\textbf {\bibinfo
  {volume} {101}},\ \bibinfo {pages} {198101} (\bibinfo {year}
  {2008})}\BibitemShut {NoStop}%
\bibitem [{\citenamefont {Giomi}\ \emph {et~al.}(2012)\citenamefont {Giomi},
  \citenamefont {Mahadevan}, \citenamefont {Chakraborty},\ and\ \citenamefont
  {Hagan}}]{Giomi2012}%
  \BibitemOpen
  \bibfield  {author} {\bibinfo {author} {\bibfnamefont {L.}~\bibnamefont
  {Giomi}}, \bibinfo {author} {\bibfnamefont {L.}~\bibnamefont {Mahadevan}},
  \bibinfo {author} {\bibfnamefont {B.}~\bibnamefont {Chakraborty}}, \ and\
  \bibinfo {author} {\bibfnamefont {M.~F.}\ \bibnamefont {Hagan}},\ }\bibfield
  {title} {\emph {\bibinfo {title} {Banding, excitability and chaos in
  active nematic suspensions},}\ }\href
  {http://stacks.iop.org/0951-7715/25/i=8/a=2245} {\bibfield  {journal}
  {\bibinfo  {journal} {Nonlinearity}\ }\textbf {\bibinfo {volume} {25}},\
  \bibinfo {pages} {2245} (\bibinfo {year} {2012})}\BibitemShut {NoStop}%
\bibitem [{\citenamefont {Edwards}\ and\ \citenamefont
  {Yeomans}(2009)}]{Edwards2009}%
  \BibitemOpen
  \bibfield  {author} {\bibinfo {author} {\bibfnamefont {S.~A.}\ \bibnamefont
  {Edwards}}\ and\ \bibinfo {author} {\bibfnamefont {J.~M.}\ \bibnamefont
  {Yeomans}},\ }\bibfield  {title} {\emph {\bibinfo {title} {Spontaneous
  flow states in active nematics: A unified picture},}\ }\href
  {http://stacks.iop.org/0295-5075/85/i=1/a=18008} {\bibfield  {journal}
  {\bibinfo  {journal} {Europhysics Letters}\ }\textbf {\bibinfo {volume}
  {85}},\ \bibinfo {pages} {18008} (\bibinfo {year} {2009})}\BibitemShut
  {NoStop}%
\bibitem [{\citenamefont {de~Gennes}\ and\ \citenamefont
  {Prost}(1993)}]{deGennes}%
  \BibitemOpen
  \bibfield  {author} {\bibinfo {author} {\bibfnamefont {P.}~\bibnamefont
  {de~Gennes}}\ and\ \bibinfo {author} {\bibfnamefont {J.}~\bibnamefont
  {Prost}},\ }\href@noop {} {\emph {\bibinfo {title} {{The Physics of Liquid
  Crystals}}}}\ (\bibinfo  {publisher} {Oxford University Press, New York},\
  \bibinfo {year} {1993})\BibitemShut {NoStop}
  \bibitem [{\citenamefont {Toner}(2012)}]{Malthus}%
  \BibitemOpen
  \bibfield  {author} {\bibinfo {author} {\bibfnamefont {J.}\ \bibnamefont
  {Toner}},\ }\bibfield  {title} {\emph {\bibinfo {title} {Birth, death, and
  flight: A theory of Malthusian flocks},}\ }\href {\doibase
  10.1103/PhysRevLett.108.088102} {\bibfield  {journal} {\bibinfo  {journal}
  {Phys. Rev. Lett.}\ }\textbf {\bibinfo {volume} {108}},\ \bibinfo {pages}
  {088102} (\bibinfo {year} {2012})}\BibitemShut {NoStop}%
\bibitem [{\citenamefont {P.M.~Chaikin}\ and\ \citenamefont
  {T.C.~Lubensky}(1995)}]{Tom's book}%
  \BibitemOpen
  \bibfield  {author} {\bibinfo {author} {\bibfnamefont {P.M.}~\bibnamefont
  {Chaikin}}\ and\ \bibinfo {author} {\bibfnamefont {T.C.}~\bibnamefont
  {Lubensky}},\ }\href@noop {} {\emph {\bibinfo {title} {{Principles of Condensed Matter Physics}}}}\ (\bibinfo  {publisher} {Cambridge University Press, Cambridge, U.K.},\
  \bibinfo {year} {1995})\BibitemShut {NoStop}
\bibitem [{\citenamefont {Thampi}\ \emph {et~al.}(2014)\citenamefont {Thampi},
  \citenamefont {Golestanian},\ and\ \citenamefont {Yeomans}}]{Thampi2014}%
  \BibitemOpen
  \bibfield  {author} {\bibinfo {author} {\bibfnamefont {S.~P.}\ \bibnamefont
  {Thampi}}, \bibinfo {author} {\bibfnamefont {R.}~\bibnamefont {Golestanian}},
  \ and\ \bibinfo {author} {\bibfnamefont {J.~M.}\ \bibnamefont {Yeomans}},\
  }\bibfield  {title} {\emph {\bibinfo {title}
  {{Vorticity, defects and correlations in active turbulence}},}\ }\href@noop {} {\bibfield  {journal} {\bibinfo
  {journal} {Philosophical Transactions of the Royal Society of London A:
  Mathematical, Physical and Engineering Sciences}\ }\textbf {\bibinfo {volume}
  {372}} (\bibinfo {year} {2014})}
  \BibitemShut {NoStop}%
  \bibitem [{\citenamefont {Leslie}(1968)}]{Leslie}%
  \BibitemOpen
  \bibfield  {author} {\bibinfo {author} {\bibfnamefont {F.~M.}\ \bibnamefont
  {Leslie}},\ }\bibfield  {title} {\emph {\bibinfo {title} {Some
  constitutive equations for liquid crystals},}\ }\href {\doibase
  10.1007/BF00251810} {\bibfield  {journal} {\bibinfo  {journal} {Archive for
  Rational Mechanics and Analysis}\ }\textbf {\bibinfo {volume} {28}},\
  \bibinfo {pages} {265--283} (\bibinfo {year} {1968})}\BibitemShut {NoStop}%
 \bibitem [{SM1()}]{SM1}%
  \BibitemOpen
  \href@noop {} {\bibinfo {title} {See Supplemental Material at [url]
  for additional information supporting the work in this paper}}\BibitemShut {NoStop}%
\bibitem [{\citenamefont {Kuzuu}\ and\ \citenamefont {Doi}(1984)}]{Doi}%
  \BibitemOpen
  \bibfield  {author} {\bibinfo {author} {\bibfnamefont {N.}\ \bibnamefont
  {Kuzuu}}\ and\ \bibinfo {author} {\bibfnamefont {M.}\ \bibnamefont
  {Doi}},\ }\bibfield  {title} {\emph {\bibinfo {title} {Constitutive
  equation for nematic liquid crystals under weak velocity gradient derived
  from a molecular kinetic equation. II. –Leslie coefficients for rodlike
  polymers–},}\ }\href {\doibase 10.1143/JPSJ.53.1031} {\bibfield  {journal}
  {\bibinfo  {journal} {Journal of the Physical Society of Japan}\ }\textbf
  {\bibinfo {volume} {53}},\ \bibinfo {pages} {1031--1038} (\bibinfo {year}
  {1984})}\BibitemShut {NoStop}%
\bibitem [{\citenamefont {Vitelli}\ and\ \citenamefont
  {Nelson}(2006)}]{Vitelli2006}%
  \BibitemOpen
  \bibfield  {author} {\bibinfo {author} {\bibfnamefont {V.}\
  \bibnamefont {Vitelli}}\ and\ \bibinfo {author} {\bibfnamefont {D.~R.}\
  \bibnamefont {Nelson}},\ }\bibfield  {title} {\emph {\bibinfo {title}
  {{Nematic textures in spherical shells}},}\ }\href@noop {} {\bibfield
  {journal} {\bibinfo  {journal} {Physical Review E}\ }\textbf {\bibinfo
  {volume} {74}},\ \bibinfo {pages} {021711} (\bibinfo {year}
  {2006})}\BibitemShut {NoStop}%
\bibitem [{\citenamefont {Lopez-Leon}\ \emph {et~al.}(2011)\citenamefont
  {Lopez-Leon}, \citenamefont {Koning}, \citenamefont {Devaiah}, \citenamefont
  {Vitelli},\ and\ \citenamefont {Fernandez-Nieves}}]{Lopez-Leon2011}%
  \BibitemOpen
  \bibfield  {author} {\bibinfo {author} {\bibfnamefont {T.}~\bibnamefont
  {Lopez-Leon}}, \bibinfo {author} {\bibfnamefont {V.}~\bibnamefont {Koning}},
  \bibinfo {author} {\bibfnamefont {K.~B.~S.}\ \bibnamefont {Devaiah}},
  \bibinfo {author} {\bibfnamefont {V.}~\bibnamefont {Vitelli}}, \ and\ \bibinfo
  {author} {\bibfnamefont {A.}~\bibnamefont {Fernandez-Nieves}},\ }\bibfield
  {title} {\emph {\bibinfo {title} {{Frustrated nematic order in spherical
  geometries}},}\ }\href@noop {} {\bibfield  {journal} {\bibinfo  {journal}
  {Nature Physics}\ }\textbf {\bibinfo {volume} {7}},\ \bibinfo {pages}
  {391--394} (\bibinfo {year} {2011})}\BibitemShut {NoStop}%
\bibitem [{\citenamefont {Fernandez-Nieves}\ \emph {et~al.}(2007)\citenamefont
  {Fernandez-Nieves}, \citenamefont {Vitelli}, \citenamefont {Utada},
  \citenamefont {Link}, \citenamefont {M\'{a}rquez}, \citenamefont {Nelson},\
  and\ \citenamefont {Weitz}}]{Fernandez-Nieves2007}%
  \BibitemOpen
  \bibfield  {author} {\bibinfo {author} {\bibfnamefont {A.}~\bibnamefont
  {Fernandez-Nieves}}, \bibinfo {author} {\bibfnamefont {V.}~\bibnamefont
  {Vitelli}}, \bibinfo {author} {\bibfnamefont {A.~S.}\ \bibnamefont {Utada}},
  \bibinfo {author} {\bibfnamefont {D.~R.}\ \bibnamefont {Link}}, \bibinfo
  {author} {\bibfnamefont {M.}~\bibnamefont {M\'{a}rquez}}, \bibinfo {author}
  {\bibfnamefont {D.~R.}\ \bibnamefont {Nelson}}, \ and\ \bibinfo {author}
  {\bibfnamefont {D.~A.}\ \bibnamefont {Weitz}},\ }\bibfield  {title} {\emph
  {\bibinfo {title} {{Novel defect structures in nematic liquid crystal
  shells.}}}\ }\href {http://www.ncbi.nlm.nih.gov/pubmed/17995213} {\bibfield
  {journal} {\bibinfo  {journal} {Phys. Rev. Lett.}\ }\textbf {\bibinfo
  {volume} {99}},\ \bibinfo {pages} {157801} (\bibinfo {year}
  {2007})}\BibitemShut {NoStop}%
\bibitem [{\citenamefont {Kamien}(2002)}]{RevModPhys.74.953}%
  \BibitemOpen
  \bibfield  {author} {\bibinfo {author} {\bibfnamefont {R.~D.}\
  \bibnamefont {Kamien}},\ }\bibfield  {title} {\emph {\bibinfo {title}
  {{The geometry of soft materials: a primer}},}\ }\href {\doibase
  10.1103/RevModPhys.74.953} {\bibfield  {journal} {\bibinfo  {journal} {Rev.
  Mod. Phys.}\ }\textbf {\bibinfo {volume} {74}},\ \bibinfo {pages} {953--971}
  (\bibinfo {year} {2002})}\BibitemShut {NoStop}%
\bibitem [{\citenamefont {Santangelo}\ \emph {et~al.}(2007)\citenamefont
  {Santangelo}, \citenamefont {Vitelli}, \citenamefont {Kamien},\ and\
  \citenamefont {Nelson}}]{Santangelo2007}%
  \BibitemOpen
  \bibfield  {author} {\bibinfo {author} {\bibfnamefont {C.~D.}\
  \bibnamefont {Santangelo}}, \bibinfo {author} {\bibfnamefont {V.}\
  \bibnamefont {Vitelli}}, \bibinfo {author} {\bibfnamefont {R.~D.}\
  \bibnamefont {Kamien}}, \ and\ \bibinfo {author} {\bibfnamefont {D.~R.}\
  \bibnamefont {Nelson}},\ }\bibfield  {title} {\emph {\bibinfo {title}
  {{Geometric theory of columnar phases on curved substrates}},}\ }\href
  {\doibase 10.1103/PhysRevLett.99.017801} {\bibfield  {journal} {\bibinfo
  {journal} {Phys. Rev. Lett.}\ }\textbf {\bibinfo {volume} {99}},\
  \bibinfo {pages} {017801-4} (\bibinfo {year} {2007})}\BibitemShut
  {NoStop}%
\bibitem [{\citenamefont {Vitelli}\ and\ \citenamefont
  {Nelson}(2004)}]{Vitelli2004}%
  \BibitemOpen
  \bibfield  {author} {\bibinfo {author} {\bibfnamefont {V.}\
  \bibnamefont {Vitelli}}\ and\ \bibinfo {author} {\bibfnamefont {D.~R.}\
  \bibnamefont {Nelson}},\ }\bibfield  {title} {\emph {\bibinfo {title}
  {Defect generation and deconfinement on corrugated topographies},}\ }\href
  {\doibase 10.1103/PhysRevE.70.051105} {\bibfield  {journal} {\bibinfo
  {journal} {Phys. Rev. E}\ }\textbf {\bibinfo {volume} {70}},\ \bibinfo
  {pages} {051105} (\bibinfo {year} {2004})}\BibitemShut {NoStop}%
\bibitem [{\citenamefont {Bowick}\ \emph {et~al.}(2004)\citenamefont {Bowick},
  \citenamefont {Nelson},\ and\ \citenamefont {Travesset}}]{Bowick2004}%
  \BibitemOpen
  \bibfield  {author} {\bibinfo {author} {\bibfnamefont {M.}\ \bibnamefont
  {Bowick}}, \bibinfo {author} {\bibfnamefont {D.~R.}\ \bibnamefont
  {Nelson}}, \ and\ \bibinfo {author} {\bibfnamefont {A.}\ \bibnamefont
  {Travesset}},\ }\bibfield  {title} {\emph {\bibinfo {title}
  {Curvature-induced defect unbinding in toroidal geometries},}\ }\href
  {\doibase 10.1103/PhysRevE.69.041102} {\bibfield  {journal} {\bibinfo
  {journal} {Phys. Rev. E}\ }\textbf {\bibinfo {volume} {69}},\ \bibinfo
  {pages} {041102} (\bibinfo {year} {2004})}\BibitemShut {NoStop}%
\bibitem [{\citenamefont {Kenyon}(1990)}]{difgeo}%
  \BibitemOpen
  \bibfield  {author} {\bibinfo {author} {\bibfnamefont {I.R.}\ \bibnamefont
  {Kenyon}},\ }\href@noop {} {\emph {\bibinfo {title} {{General Relativity}}}}\
  (\bibinfo  {publisher} {Oxford Science Publications},\ \bibinfo {year}
  {1990})\BibitemShut {NoStop}%
\bibitem [{\citenamefont {Pairam}\ \emph {et~al.}(2013)\citenamefont {Pairam},
  \citenamefont {Vallamkondu}, \citenamefont {Koning}, \citenamefont {van
  Zuiden}, \citenamefont {Ellis}, \citenamefont {Bates}, \citenamefont
  {Vitelli},\ and\ \citenamefont {Fernandez-Nieves}}]{Pairam2013}%
  \BibitemOpen
  \bibfield  {author} {\bibinfo {author} {\bibfnamefont {E.}\ \bibnamefont
  {Pairam}}, \bibinfo {author} {\bibfnamefont {J.}\ \bibnamefont
  {Vallamkondu}}, \bibinfo {author} {\bibfnamefont {V.}\ \bibnamefont
  {Koning}}, \bibinfo {author} {\bibfnamefont {B.~C.}\ \bibnamefont {van
  Zuiden}}, \bibinfo {author} {\bibfnamefont {P.~W.}\ \bibnamefont {Ellis}},
  \bibinfo {author} {\bibfnamefont {M.~A.}\ \bibnamefont {Bates}}, \bibinfo
  {author} {\bibfnamefont {V.}\ \bibnamefont {Vitelli}}, \ and\ \bibinfo
  {author} {\bibfnamefont {A.}\ \bibnamefont {Fernandez-Nieves}},\
  }\bibfield  {title} {\emph {\bibinfo {title} {{Stable nematic droplets
  with handles}},}\ }\href {\doibase 10.1073/pnas.1221380110} {\bibfield
  {journal} {\bibinfo  {journal} {Proceedings of the National Academy of
  Sciences}\ }\textbf {\bibinfo {volume} {110}},\ \bibinfo {pages} {9295--9300}
  (\bibinfo {year} {2013})}\BibitemShut {NoStop}%
\bibitem [{\citenamefont {Koning}\ \emph {et~al.}(2014)\citenamefont {Koning},
  \citenamefont {van Zuiden}, \citenamefont {Kamien},\ and\ \citenamefont
  {Vitelli}}]{Koning2014}%
  \BibitemOpen
  \bibfield  {author} {\bibinfo {author} {\bibfnamefont {V.}\ \bibnamefont
  {Koning}}, \bibinfo {author} {\bibfnamefont {B.~C.}\ \bibnamefont {van
  Zuiden}}, \bibinfo {author} {\bibfnamefont {R.~D.}\ \bibnamefont
  {Kamien}}, \ and\ \bibinfo {author} {\bibfnamefont {V.}\ \bibnamefont
  {Vitelli}},\ }\bibfield  {title} {\emph {\bibinfo {title} {{Saddle-splay
  screening and chiral symmetry breaking in toroidal nematics}},}\ }\href
  {\doibase 10.1039/C4SM00076E} {\bibfield  {journal} {\bibinfo  {journal}
  {Soft Matter}\ }\textbf {\bibinfo {volume} {10}},\ \bibinfo {pages}
  {4192--4198} (\bibinfo {year} {2014})}\BibitemShut {NoStop}%
\bibitem [{\citenamefont {Davidson}\ \emph {et~al.}(2015)\citenamefont
  {Davidson}, \citenamefont {Kang}, \citenamefont {Jeong}, \citenamefont
  {Still}, \citenamefont {Collings}, \citenamefont {Lubensky},\ and\
  \citenamefont {Yodh}}]{Davidson2015}%
  \BibitemOpen
  \bibfield  {author} {\bibinfo {author} {\bibfnamefont {Z.~S.}\ \bibnamefont
  {Davidson}}, \bibinfo {author} {\bibfnamefont {L.}\ \bibnamefont {Kang}},
  \bibinfo {author} {\bibfnamefont {J.}\ \bibnamefont {Jeong}}, \bibinfo
  {author} {\bibfnamefont {T.}\ \bibnamefont {Still}}, \bibinfo {author}
  {\bibfnamefont {P.~J.}\ \bibnamefont {Collings}}, \bibinfo {author}
  {\bibfnamefont {T.~C.}\ \bibnamefont {Lubensky}}, \ and\ \bibinfo {author}
  {\bibfnamefont {A.~G.}\ \bibnamefont {Yodh}},\ }\bibfield  {title} {\emph
  {\bibinfo {title} {{Chiral structures and defects of lyotropic chromonic
  liquid crystals induced by saddle-splay elasticity}},}\ }\href {\doibase
  10.1103/PhysRevE.91.050501} {\bibfield  {journal} {\bibinfo  {journal}
  {Physical Review E}\ }\textbf {\bibinfo {volume} {91}},\ \bibinfo {pages}
  {050501} (\bibinfo {year} {2015})}\BibitemShut {NoStop}%
\bibitem [{\citenamefont {Sokolov}\ \emph {et~al.}(2015)\citenamefont
  {Sokolov}, \citenamefont {Zhou}, \citenamefont {Lavrentovich},\ and\
  \citenamefont {Aranson}}]{Sokolov2015}%
  \BibitemOpen
  \bibfield  {author} {\bibinfo {author} {\bibfnamefont {A.}\ \bibnamefont
  {Sokolov}}, \bibinfo {author} {\bibfnamefont {S.}\ \bibnamefont {Zhou}},
  \bibinfo {author} {\bibfnamefont {O.~D.}\ \bibnamefont {Lavrentovich}}, \
  and\ \bibinfo {author} {\bibfnamefont {I.~S.}\ \bibnamefont {Aranson}},\
  }\bibfield  {title} {\emph {\bibinfo {title} {Individual behavior and
  pairwise interactions between microswimmers in anisotropic liquid},}\ }\href
  {\doibase 10.1103/PhysRevE.91.013009} {\bibfield  {journal} {\bibinfo
  {journal} {Phys. Rev. E}\ }\textbf {\bibinfo {volume} {91}},\ \bibinfo
  {pages} {013009} (\bibinfo {year} {2015})}\BibitemShut {NoStop}%
\bibitem [{\citenamefont {Guo}\ \emph {et~al.}(2016)\citenamefont {Guo},
  \citenamefont {Jiang}, \citenamefont {Peng}, \citenamefont {Sun},
  \citenamefont {Yaroshchuk}, \citenamefont {Lavrentovich},\ and\ \citenamefont
  {Wei}}]{Guo2016}%
  \BibitemOpen
  \bibfield  {author} {\bibinfo {author} {\bibfnamefont {Y.}\ \bibnamefont
  {Guo}}, \bibinfo {author} {\bibfnamefont {M.}\ \bibnamefont {Jiang}},
  \bibinfo {author} {\bibfnamefont {C.}\ \bibnamefont {Peng}}, \bibinfo
  {author} {\bibfnamefont {K.}\ \bibnamefont {Sun}}, \bibinfo {author}
  {\bibfnamefont {O.}\ \bibnamefont {Yaroshchuk}}, \bibinfo {author}
  {\bibfnamefont {O.}\ \bibnamefont {Lavrentovich}}, \ and\ \bibinfo {author}
  {\bibfnamefont {Q.H.}\ \bibnamefont {Wei}},\ }\bibfield  {title} {\emph
  {\bibinfo {title} {High-resolution and high-throughput plasmonic
  photopatterning of complex molecular orientations in liquid crystals},}\
  }\href {\doibase 10.1002/adma.201506002} {\bibfield  {journal} {\bibinfo
  {journal} {Advanced Materials}\ }\textbf {\bibinfo {volume} {28}},\ \bibinfo
  {pages} {2353--2358} (\bibinfo {year} {2016})}\BibitemShut {NoStop}%
 \bibitem [{SM()}]{SM}%
  \BibitemOpen
  \href@noop {} {\bibinfo {title} {See Supplemental Material at [url]
  for movies of flows in selected geometries}}\BibitemShut {NoStop}%
\bibitem [{\citenamefont {Mermin}\ and\ \citenamefont {Wagner}(1966)}]{MW}%
  \BibitemOpen
  \bibfield  {author} {\bibinfo {author} {\bibfnamefont {N.~D.}\ \bibnamefont
  {Mermin}}\ and\ \bibinfo {author} {\bibfnamefont {H.}~\bibnamefont
  {Wagner}},\ }\bibfield  {title} {\emph {\bibinfo {title} {Absence of
  ferromagnetism or antiferromagnetism in one- or two-dimensional isotropic
  heisenberg models},}\ }\href {\doibase 10.1103/PhysRevLett.17.1133}
  {\bibfield  {journal} {\bibinfo  {journal} {Phys. Rev. Lett.}\ }\textbf
  {\bibinfo {volume} {17}},\ \bibinfo {pages} {1133--1136} (\bibinfo {year}
  {1966})}\BibitemShut {NoStop}%
\bibitem [{\citenamefont {Hohenberg}(1967)}]{MW'}%
  \BibitemOpen
  \bibfield  {author} {\bibinfo {author} {\bibfnamefont {P.~C.}\ \bibnamefont
  {Hohenberg}},\ }\bibfield  {title} {\emph {\bibinfo {title} {Existence of
  long-range order in one and two dimensions},}\ }\href {\doibase
  10.1103/PhysRev.158.383} {\bibfield  {journal} {\bibinfo  {journal} {Phys.
  Rev.}\ }\textbf {\bibinfo {volume} {158}},\ \bibinfo {pages} {383--386}
  (\bibinfo {year} {1967})}\BibitemShut {NoStop}%
\bibitem [{\citenamefont {Crawford}\ \emph {et~al.}(1991)\citenamefont
  {Crawford}, \citenamefont {Allender},\ and\ \citenamefont
  {Doane}}]{Crawford1991}%
  \BibitemOpen
  \bibfield  {author} {\bibinfo {author} {\bibfnamefont {G.~P.}\ \bibnamefont
  {Crawford}}, \bibinfo {author} {\bibfnamefont {D.~W.}\ \bibnamefont
  {Allender}}, \ \bibinfo {author} {\bibfnamefont {J.~W.}\ \bibnamefont
  {Doane}}, \bibinfo {author} {\bibfnamefont {M}\ \bibnamefont
  {Vilfan}}, and\ \bibinfo {author} {\bibfnamefont {I.}\ \bibnamefont
  {Vilfan}}\ }\bibfield  {title} {\emph {\bibinfo {title} {{Finite molecular
  anchoring in the escaped-radial nematic configuration}},}\ }\href@noop {}
  {\bibfield  {journal} {\bibinfo  {journal} {Physical Review A}\ }\textbf
  {\bibinfo {volume} {44}},\ \bibinfo {pages} {2570--2577} (\bibinfo {year} {1991})}\BibitemShut {NoStop}%
  \bibitem [{\citenamefont {Bates}(1997)}]{Bates1997}%
  \BibitemOpen
  \bibfield  {author} {\bibinfo {author} {\bibfnamefont {J.~W.}\ \bibnamefont
  {Bates}},\ }\bibfield  {title} {\emph {\bibinfo {title} {{On toroidal Green's functions}},}\ }\href {\doibase 10.1063/1.532061} {\bibfield  {journal} {\bibinfo
   {journal} {Journal of Mathematical Physics}\ }\textbf {\bibinfo {volume}
  {38}},\ \bibinfo {pages} {3679} (\bibinfo {year} {1997})}\BibitemShut
  {NoStop}%
\bibitem [{\citenamefont {{A. Rotenberg}}(1960)}]{ARotenberg1959}%
  \BibitemOpen
  \bibfield  {author} {\bibinfo {author} {\bibnamefont {{A Rotenberg}}},\
  }\bibfield  {title} {\emph {\bibinfo {title} {{The calculation of toroidal harmonics}},}\ }\href@noop {} {\bibfield  {journal} {\bibinfo  {journal} {Math. Comp.}\
  }\textbf {\bibinfo {volume} {14}},\ \bibinfo {pages} {274} (\bibinfo {year}
  {1960})}\BibitemShut {NoStop}%
\end{thebibliography}

%

\end{document}



\title{The geometry of thresholdless active flow in nematic microfluidics - Supplemental Material}

\author{Richard Green}
\affiliation{Instituut-Lorentz, Universiteit Leiden, 2300 RA Leiden, The Netherlands}
\author{John Toner}
\affiliation{Department of Physics and Institute of Theoretical
Science, University of Oregon, Eugene, OR $97403^1$}
\affiliation{Max Planck Institute for the Physics of Complex Systems, N\"othnitzer Str. 38, 01187 Dresden, Germany}
\author{Vincenzo Vitelli}
\affiliation{Instituut-Lorentz, Universiteit Leiden, 2300 RA Leiden, The Netherlands}

\maketitle

In this Supplemental Material, we prove  a number of claims made in the main text \cite{Main}, and give the details of some of the calculations summarized therein. In particular, in section I, we demonstrate the applicability of the hydrodynamic equations we use here, suitably modified to take into account the presence of multiple components, to ``living liquid crystals". In section II, we show that non-equilibrium molecular fields are negligible. In section III we justify the frozen director approximation, and the isotropic viscosity approximation, in the limit of weak nematic order. This section also includes a  demonstration of the irrelevance of nematic order parameter amplitude fluctuations in all cases. Section IV presents the technical mathematical details of our derivation of the geometric integral conditions for thresholdless active flow, while in section V we give the derivation of a condition on the vorticity for flow in the ``frozen director" regime. In sections VI and VII we respectively give the details of our analysis of the Frederiks cell active pump, and active flow in a thin toroidal shell.

\section{\uppercase{Living Liquid Crystals}}

These systems are a mixture of living bacteria, which provide the activity, and a background medium composed of nematically ordered non-active molecules. That such a system is an active nematic can be seen on symmetry grounds: the symmetry is that of a nematic, and the bacteria are active. Hence, the composite system as a whole {\it is} an active nematic and since the hydrodynamic equations are, as always, determined by symmetry (and  the fact that the system is active), the hydrodynamic equations will be those we've considered here, except for one detail: living liquid crystals are a {\it mixture} of two components (i.e., bacteria, and background liquid crystal), and each component is separately conserved (assuming we neglect birth and death of the bacteria, which is frequently but not always the case; see \cite{Malthus}). Hence, there is one more conserved variable in this two-component active nematic and, as usual in hydrodynamics \cite{Tom's book}, conserved quantities are hydrodynamic variables. So the two-component active nematic has one more hydrodynamic variable than the one-component active nematic that we've considered throughout the rest of this paper, which, of course, changes the hydrodynamics. We can take this extra hydrodynamic variable to be the concentration $c(\bf{r}, t)$ of bacteria. The detailed effects of this modification of the hydrodynamics will be addressed in a future publication. Here, we limit ourselves to the observation, which will be proven in that future publication, that the hydrodynamic equations are sufficiently similar in that case that the criterion $\nabla \times \bf{f_a}\ne \bf{0}$ for thresholdless flow, described in detail after Eq. (1) of the main text, is unchanged.

\section{\uppercase{Irrelevance of non-equilibrium molecular fields}}

As noted in the main text \cite{Main}, the equations of motion we have used are not, in fact, the most general. In particular, the  Frank free energy $F$ 
appearing in the Navier-Stokes Eq. (\vEOM) below need not, and, indeed, will not be equal to that 
in the
nematodynamic Eq. (\nEOM), since we are dealing with  a non-equilibrium system. Distinguishing these two free energies by writing them as $F_v$ and $F_n$, we will now demonstrate that the fact that $F_v\ne F_n$ affects neither our criterion for flow, nor, for small activity, the flow that occurs when this criterion is met.

Rewriting the equations of motion (\vEOM-\continuity) taking into account this difference gives:
\begin{subequations}
\begin{eqnarray}
\rho_0 \frac{D v_k}{D t} &=& - \partial_k P + \eta \nabla^2 v_k + \alpha \partial_j \left(n_j n_k  \right)  
+\partial_j(\lambda_{ijk}{h_{vi}})
 \nonumber \\
  \label{eq: SvEOM} \\ 
\frac{D n_i}{D t}
&=& \lambda_{ijk} \partial_jv_k-
 \frac{1}{\gamma_1} \left[h_{ni} - \left(\bm{h}_n \cdot \hat{\bm{n}} \right) n_i\right]
\label{eq: SnEOM} \\
\nabla \cdot \bm{v} &=& 0, \label{eq: Scontinuity}
 \end{eqnarray}
 \end{subequations}
where the  ``molecular fields" $\bm{h}_\nu$, $\nu=[v,n]$  appearing in these equations are given by $\bm{h}_\nu=\frac{\delta F_\nu}{\delta \nh}$, 
which implies
\begin{eqnarray}
\bm{h}_\nu&\equiv&\frac{\delta F_\nu}{\delta \nh}= 2(K_{2\nu}-K_{3\nu})\left(\hat{\bm{n}} \cdot \left( \nabla \times \hat{\bm{n}} \right)\right)  \nabla \times \hat{\bm{n}} \nonumber\\ &-&K_{3\nu}\nabla^2\hat{\bm{n}}
+ (K_{3\nu}-K_{2\nu}) \hat{\bm{n}} \times \nabla \left(\hat{\bm{n}}\cdot \nabla \times \hat{\bm{n}} \right)\nonumber\\ &+&(K_{3\nu}-K_{1\nu})\nabla\left(\nabla\cdot\hat{\bm{n}}\right).
\label{hexp}
\end{eqnarray}
Here the 
 the $F_\nu$'s, $\nu=[v,n]$ are the non-equilibrium generalizations of the equilibrium Frank free energy $F$. They are constrained by rotation invariance in exactly the same way as in equilibrium, and must, therefore, both take the usual Frank free energy \cite{deGennes} {\it form}: 
\begin{eqnarray}
F_\nu &=& \frac{1}{2}\int d^3r[K_{1\nu} \left(\nabla\cdot\hat{\bm{n}}\right)^2+ K_{2\nu}\left(\hat{\bm{n}}\cdot \left( \nabla \times \hat{\bm{n}} \right)\right)^2 \nonumber \\
&& \>\>\>\>\>\>\>\>\>\>\>\>\>\>\>\> + K_{3\nu}\left|\hat{\bm{n}}\times \left( \nabla \times \hat{\bm{n}} \right)\right|^2 ]\,,
\label{Frank FE SM}
\end{eqnarray}
However, although the {\it form} of the two free energies must be the same, away from equilibrium, the values of the Frank constants $K_{1,2,3}$ need not be the same in the two free energies. Only 
in equilibrium, in which the activity parameter $\alpha=0$, do the two Frank free energies become equal  ($F_v=F_n$). 
In an active system, however, the fundamentally non-equilibrium nature
of the problem means that there are no such requirements of equality; that is,  $F_v\ne F_n$ away from equilibrium, in contrast to the equations of motion (2a-2c) of the main text, in which we took $F_v= F_n=F$. Our point here is that this is not, strictly speaking, true. Nonetheless, 
we do expect \cite{actnem} that both $\alpha$ and the difference between $F_v$ and $F_n$ will be proportional to the density of active particles, and will, hence, be very small when that density is small. Since we are interested in the small activity (i.e., low active particle density) limit, in the main text we ignore the difference between $F_v$ and  $F_n$. In this section, we show that
none of the conclusions of the main text are altered by this difference. Specifically, we demonstrate first, that for small activity, none of the flow fields $\bm{v}(\bm{r})$ that we calculated are quantitatively affected to leading order in activity $\alpha$. Then we will show that our conclusions about the classes of configurations that do not exhibit thresholdless flow (e.g., pure twist) are completely unaffected by these differences to {\it any} order in $\alpha$. This justifies our neglect of those differences in the main text.

We begin by considering situations in which the activity $\alpha$ {\it does} induces thresholdless flow. We will show that, even when $\alpha$ is very small, the terms involving $\hvv$ in (\vEOMtwo) are {\it always} 
negligible, relative to the $\alpha$ terms. This is because, in the small activity limit, $\hvv\rightarrow\hvn$, with the difference $\hvv-\hvn\propto \alpha$, since, as noted earlier, the difference between $\hvv$ and $\hvn$ is a purely active effect. However, we have already shown in the main text that $\hvn\parallel\nh$; it is straightforward to show that when $\hvv\parallel\nh$, the terms involving $\hvv$ in (\vEOMtwo) vanish. Hence, the only piece of those terms that can survive must arise from the {\it difference}  $\hvv-\hvn$, which, as we've just argued, is proportional to $\alpha$. However, these terms also involve more spatial derivatives of $\nh$ than the $\alpha$ term (to be precise, three versus one) and so on dimensional grounds, we expect the ratio of the $\hvv$ terms to the $\alpha$ terms to be $\sim\left( \frac{a}{L}\right)^2$,  where $a$ is a microscopic length (e.g., the size of the active particles), while $L$ is the length scale over which $\nh$ varies (usually a macroscopic length). Hence, the $\hvv$ terms in (\vEOMtwo) are completely negligible, regardless of the value of $\alpha$, in a macroscopic geometry. 
Note that the length $a$ that appears in this estimate can {\it not} be the instability length
$ L_{\text{inst}}\sim\sqrt{\frac{K}{\alpha}}$ discussed in the main text, since, as we just argued, the ratio of the $\hvv$ to the $\alpha$ term must be independent of $\alpha$.
 
For the second case, in which there is no thresholdless flow, the nematic director must minimize the Frank free energy $F_n$. The Euler-Lagrange equations for the director field configurations $\bm{{\hat n}}_{\nu}(\bm{r})$ minimising $F_{\nu}$ are then simply
\begin{equation}
\bm{h}_{\nu} = \mu_{\nu} (\bm{r}) \bm{n}_{\nu} \,,\label{ELG2}
\end{equation}
\noindent where $\mu_\nu (\bm{r})$ is a Lagrange multiplier. 
The active force $\bm{f}_a$ may be rewritten
\begin{equation}
\label{eq: active force SM}
\bm{f}_a = \alpha \left[ \hat{\bm{n}} \> \nabla \cdot \hat{\bm{n}} - \hat{\bm{n}} \times \left( \nabla \times \hat{\bm{n}} \right) \right] \,.
\end{equation}

Now let's consider the cases in which we argued in the main text that  there could be no thresholdless flow. We'll start with the case in which a pure twist configuration (i.e., one with $ \nabla \cdot \hat{\bm{n}}=0$ and  $\hat{\bm{n}} \times \left( \nabla \times \hat{\bm{n}}\right)=\bm{0}$ of $\bm{{\hat n}}(\bm{r})$) minimizes $F_n$, the free energy appearing in the $\bm{{\hat n}}(\bm{r})$ equation of motion.
We can immediately see that  such a director field  has no active force.
One might wonder whether active flow in this case can be induced by the activity-induced difference between $\hvv$ 
and $\hvn$; we'll now prove that this is not the case. 

To see this, note that in a pure twist state, since $\nh\times(\nabla\times \nh)=\bm{0}$, $\nabla\times \nh$ must be parallel to $\nh$ itself. This implies
\begin{equation}
\nabla\times \nh=g(\rv)\nh(\rv)
\label{tw1}
\end{equation}
where $g(\rv)$ is some scalar function of $\rv$. Furthermore, since $\nh$ is divergenceless in a pure twist state ($\nabla\cdot\nh=0$),
a well-known identity of vector calculus implies $\nabla^2\nh=-\nabla\times(\nabla\times\nh)$; using (\ref{tw1}) in this identity gives
\begin{equation}
\nabla^2 \nh=-g\nabla\times\nh-\nabla g\times\nh=-g^2\nh+\nh\times\nabla g \,,
\label{lap}
\end{equation}
where in the second equality we have used (\ref{tw1}) a second time.
Using (\ref{tw1}), (\ref{lap}) and $\nabla\cdot\nh=0$ in our expression (\ref{hexp}) for the molecular field $\hvn$ gives 
\begin{equation}
\hvn=(2K_{2n}-K_{3n})g^2 \nh-K_{2n}\nh\times\nabla  g \,.
\label{hn}
\end{equation}
 A moment's reflection reveals that, for this to be parallel to $\nh$, as it must be if we are to satisfy the director equation of motion (\nEOMtwo) with $\bm{v}=\bm{0}$, we must have $\nabla g \parallel \nh$. For such a $g$, (\ref{hn}) implies
\begin{equation}
\hvn=(2K_{2n}-K_{3n})g^2 \nh \,,
\end{equation}
and, by the same reasoning, 
\begin{equation}
\hvv=(2K_{2v}-K_{3v})g^2 \nh \,.
\label{hv}
\end{equation}
Thus, for any pure twist configuration that gives $\hvn\propto\nh$ (which is just the condition for minimizing the Frank energy $F_n$ subject to the constraint $|\nh|=1$),  the active force $\bm{f}_a$ vanishes, and both $\hvv$ and $\hvn$ are everywhere parallel to $\nh$. But the latter conditions imply, as noted earlier, that all of the terms involving   $\hvv$ and $\hvn$
in (\vEOMtwo) and (\nEOMtwo) vanish. Since $\fa$ does as well, and all of the other terms in those equations vanish when $\vv=\bm{0}$, we can conclude that, if $\nh$ is in a pure twist configuration that minimizes $F_n$, there will be no thresholdless active flow.

\section{\uppercase{Justification of the frozen director and isotropic viscosity tensor approximations and application in the case of weak nematic order }}

We demonstrate here that the condition $\gamma_1\ll\eta$ is sufficient to justify the frozen director approximation. We also show that any system with weak nematic order will be in this limit.  We begin by noting that if the active and viscous terms are balanced in Eq. (\vEOM), this implies schematically that if the system has a characteristic length scale $L$, then $\eta v / L^2 \sim \alpha / L$ and so $v \sim \alpha L/\eta$. This last result implies that the Reynolds number $Re\equiv\frac{\rho_0 v  L}{\eta}=\frac{\rho_0 L^2 \alpha}{\eta^2}$. Furthermore, using this estimate of $v$ in Eq. (\nEOM), we obtain 
\begin{equation}
\frac{1}{\gamma_1}\frac{\delta F}{\delta \bm{\hat{n}}} \sim \alpha/\eta .
\label{hest}
\end{equation}

Assuming that $\alpha$ is small enough that $Re \ll 1$, we can make the familiar Stokes approximation of neglecting the inertial terms on the left hand side of Eq. (\vEOM). Finally, we may neglect the $\lambda$ term in Eq. (\vEOM), which is of order $\frac{\gamma_1}{\eta} \frac{\alpha}{L}$ , and is therefore smaller than the unperturbed active force by a factor of $\frac{\gamma_1}{\eta}$. 

We also need to take into account the change in the active force resulting from the change in the director field $\delta \bm{\hat{n}}\equiv\bm{\hat{n}}-\bm{\hat{n}}_0$ induced by the flow; here $\bm{\hat{n}}_0$ is the equilibrium configuration of the nematic director (that is, the one that minimizes the Frank free energy, or, equivalently, the field that is present before the activity is switched on). Since schematically the molecular field $\frac{\delta F}{\delta \bm{\hat{n}}} \sim \frac{K \delta n}{L^2}$ (note that $\delta n$ appears in this expression rather than $n$ because $\left(\frac{\delta F}{\delta \bm{\hat{n}}}\right)_{\bm{\hat{n}}=\bm{\hat{n}}_0}=\bm{0}$), 
our estimate (\ref{hest}) of that field implies that the magnitude $\delta n$ of the perturbation in the director field  must be of order $\frac{\alpha \gamma_1 L^2}{\eta K}\sim\frac{\gamma_1}{\eta} \left(\frac{L}{L_{\text{inst}}}\right)^2$, where $L_{\text{inst}}$ is the length-scale beyond which the uniform state becomes unstable. 
Since we are considering systems which are smaller than this length, and since we are also 
assuming $\gamma_1 \ll \eta$, the change $\delta n$ in $\bf{\hat{n}}$ is $\ll \bf{\hat{n}}_0$, the undistorted director configuration, and, hence, negligible.

To summarize: in the  ``frozen director" regime, defined as $\gamma_1 \ll \eta$,  and small activity $\alpha \ll K/L^2$,  we can determine the flow field simply by balancing the viscous force $\eta\nabla^2 \bf{v}$ plus the pressure gradient $\nabla  P$ against the active force $\fa$ computed for the unperturbed, equilibrium configuration $\bn_0$ which minimizes the Frank free energy; that is, we can take the active force 
\begin{eqnarray}
\fa=\alpha \left( \bm{{\hat n}} \cdot \nabla \bm{{\hat n}} + \bm{{\hat n}}\nabla \cdot \bm{{\hat n}}\right)\approx\alpha \left( \bm{{\hat n}}_0 \cdot \nabla \bm{{\hat n}}_0 + \bm{{\hat n}}_0\nabla \cdot \bm{{\hat n}}_0\right)\,.\nonumber\\
\label{frozen}
\end{eqnarray}

Making this substitution, and neglecting the $\lambda$-term in (\vEOM), simplifies (\vEOM)-(\continuity) to:

\begin{eqnarray}
\label{eqn: simplified Navier Stokes}
0 &=& - \nabla P + \eta \nabla^2 \bm{v} + \alpha \left( \bm{{\hat n_0}} \cdot \nabla \bm{{\hat n_0}} + \bm{{\hat n_0}}\nabla \cdot \bm{{\hat n_0}}\right) 
 \end{eqnarray}

\noindent with $\nabla \cdot \bm{v} =0$. 

We now justify the isotropic viscosity approximation. In the limit of weak order, which in the notation of Kuzuu and Doi \cite{Doi} is the limit $S_2, S_4 \ll1$,  our isotropic viscosity approximation becomes valid because $\alpha_4$ (our activity parameter $\alpha$ should not be confused with Kuzuu and Doi 's anisotropic viscosities), which is just the isotropic piece of the viscosity, is much greater than the anisotropic pieces $\alpha_{1, 5, 6}$ of the viscosity, since the latter all vanish when $S_{2,4}\rightarrow0$) with (again in the notation of Kuzuu and Doi), $\eta=\alpha_4 /2 \approx \eta^* C^3 r^2$. Furthermore, the coefficient $\gamma_1=\alpha_3-\alpha_2=10 \eta^* C^3 r^2 S_2/\lambda$. Taking the ratio $\gamma_1/\eta$ then gives $\gamma_1/\eta \approx  10 S_2/\lambda$, which, is always much less than $1$ when the order is weak, since the flow alignment parameter $\lambda$ is typically $O(1)$, and $S_2 \ll1$ when the order is weak. 

We therefore expect our analytic solutions for the velocity fields, which assumed both isotropic viscosity and $\gamma_1\ll\eta$ (to justify the ``frozen director"
approximation) to be quantitatively accurate in all active systems in which the nematic order is weak.

Note that no matter how strong the order is, at sufficiently long wavelengths, fluctuations in the nematic order parameter are much smaller than fluctuations in the nematic director. Hence, the director field representation is always a good approximation at sufficiently long wavelengths. In the case of weak nematic order, the nematic correlation length is of order $a/S$ where $a$ is a molecular length ($\sim 1$nm) and $S$ is the nematic order parameter. Taking $S \sim 0.01$, the nematic correlation length is of order $100$nm, much smaller than the size of a millimeter-sized sample.

We also note that this  in particular implies that the ``frozen director" approximation will always be valid for systems  close to a weakly first order nematic to isotropic (NI) transition; since many NI transitions are, indeed, weakly first order \cite{deGennes}, this means it should be quite easy to experimentally test our quantitative predictions for the flow field.

Next, in the remainder of this section we extend the results of section II in the case of a \emph{simply connected} sample in the ``frozen director" approximation, in which case the condition for thresholdless flow $\nabla \times \bm{f_a} \ne \bm{0}$ is necessary as well as sufficient. We consider the case in which the director field that minimizes $F_n$ is  pure splay, by which we mean $\nabla \times \hat{\bm{n}}=\bm{0}$. The curl of the active force is now given by
\begin{eqnarray}
\nabla \times \bm{f}_a &=& \alpha \nabla \times \left[ \hat{\bm{n}} \> \nabla \cdot \hat{\bm{n}} \right] \nonumber \\
&=& \alpha \nabla \left(\nabla \cdot \hat{\bm{n}} \right) \times \hat{\bm{n}},
\end{eqnarray}
\noindent which is also zero when the pure splay director field is a ground state of $F_n$, because the Euler-Lagrange equations that arise from minimizing $F_n$ then require that  $h_n\parallel \nh$, which in turn, from (\ref{hexp}), requires that $\nabla \left(\nabla \cdot \hat{\bm{n}} \right)$ is parallel to $\hat{\bm{n}}$. Furthermore, when 
$\nabla \times \hat{\bm{n}}=\bm{0}$, we can write $\nh=\nabla\Phi(\rv)$, which then implies $\hvv=-K_{1v}\nabla^2\nabla\Phi$, and $\hvn=-K_{1n}\nabla^2\nabla\Phi$. Thus, $\hvv\parallel\hvn$, so, if $\hvn\parallel\nh$ everywhere, $\hvv\parallel\nh$ everywhere as well. Hence, once again, the $\hvv$ and $\hvn$ terms in (\vEOMtwo) and (\nEOMtwo), respectively, vanish, as does the curl of the active force. Under the conditions of this section we can conclude that there is no flow. 

We now turn to the case of a pure bend field (i.e., one for which $\nabla \cdot \bm{\hat{n}} = \bm{\hat{n}} \cdot \nabla \times \bm{\hat{n}} =0$). Using the identity $\nabla(\bm{A} \cdot \bm{B})= (\bm{A} \cdot \nabla) \bm{B} + (\bm{B} \cdot \nabla) \bm{A} + \bm{A} \times (\nabla \times \bm{B}) + \bm{B} \times (\nabla \times \bm{A})$, with  $\bm{A}=\bm{\hat{n}}$ and $\bm{B}=  \nabla \times \bm{\hat{n}}$, and recalling that $\bm{\hat{n}} \cdot \nabla \times \bm{\hat{n}}$ must vanish in a pure bend field, gives:
\begin{equation}
 (\bm{\hat{n}} \cdot \nabla) \nabla \times \bm{\hat{n}} + (\nabla \times \bm{\hat{n}} \cdot \nabla) \bm{\hat{n}} + \bm{\hat{n}} \times (\nabla \times \nabla \times \bm{\hat{n}}) = 0.
 \label{curl1}
\end{equation}
 If this pure bend state is also a ground state of $F_n$, the Euler-Lagrange Eq. (\ref{ELG2}) for $F_n$ is satisfied. For pure bend, that equation  reduces to $\nabla^2 \bm{\hat{n}} \parallel \bm{\hat{n}}$, so that $\bm{\hat{n}} \times (\nabla \times \nabla \times \bm{\hat{n}}) $  $ = - \bm{\hat{n}} \times \nabla^2 \bm{\hat{n}}$ $= \bf{0}$, thereby eliminating the last term of (\ref{curl1}).
To compute $\nabla \times \bm{f}_a$, we now use the identity $ \nabla \times (\bm{A} \times \bm{B}) = \bm{A} (\nabla \cdot \bm{B}) - \bm{B} (\nabla \cdot \bm{A}) + (\bm{B} \cdot \nabla) \bm{A} - (\bm{A} \cdot \nabla) \bm{B}$, again with  $\bm{A}=\bm{\hat{n}}$ and $\bm{B}=  \nabla \times \bm{\hat{n}}$, for $-\bm{\hat{n}} \times (\nabla \times \bm{\hat{n}})$, to get
\begin{equation}
 \nabla \times \bm{f}_a = (\bm{\hat{n}} \cdot \nabla) \nabla \times \bm{\hat{n}} - (\nabla \times \bm{\hat{n}} \cdot \nabla) \bm{\hat{n}}. 
\end{equation} 
\noindent Using our previous result (\ref{curl1}) together with $\bm{\hat{n}} \times (\nabla \times \nabla \times \bm{\hat{n}}) = \bf{0}$, we can rewrite this equation as
\begin{equation}
 \nabla \times \bm{f}_a = 2 (\bm{\hat{n}} \cdot \nabla) \nabla \times \bm{\hat{n}} = -2 (\nabla \times \bm{\hat{n}} \cdot \nabla) \bm{\hat{n}}.
 \label{purebend}
\end{equation}

This is as far as we can go considering completely general pure bend configurations. To proceed further, we will now, in addition to imposing pure bend, add the additional restriction to ``2D" configurations, by which we mean that $\bm{\hat{n}}$  only depends on $x$ and $y$, and has no $z$ component, in some Cartesian coordinate system. Then  $\nabla \times \bm{\hat{n}}$ is in the $z$-direction, and so $(\nabla \times \bm{\hat{n}} \cdot \nabla) \bm{\hat{n}}=0$,  which implies from (\ref{purebend}) that  $\nabla \times \bm{f}_a=\bf{0}$ as well.
Similarly we have that $\hvn = -K_{3n} \nabla^2 \nh \parallel \nh$ by virtue of the Euler-Lagrange equations. Since $\hvv = -K_{3v} \nabla^2 \nh$, this is also parallel to $\nh$ and so the $\hvv$ terms vanish, contributing nothing to Eq. (\vEOMtwo).

We can thus conclude that under the conditions of this section, a two-dimensional active nematic with a director field in its ground state must have \emph{both splay and bend} for there to be thresholdless flow in the absence of external fields.

For fully three-dimensional configurations of an active nematic, on the other hand, for which there is also twist to take into account, it is unclear whether or not both splay and bend are necessary for thresholdless flow to occur in the absence of external fields. 

What we can conclude though is that, under the conditions of this section and in the absence of external fields, the ground state director field must at the very least either have both splay and twist,  or have bend, in order to induce thresholdless active flow.

In summary, we have identified three large classes of spatially non-uniform director configurations, namely, {\it all} pure twist and in the case of simply connected geometries in the ``frozen director" approximation, all pure splay, and pure 2D bend, which do not induce thresholdless active flow. Thus, the requirements for thresholdless active flow are far more stringent than the mere existence of a spatially non-uniform director field.

\section{\uppercase{Derivation of geometric integral conditions for thresholdless active flow}}
\begin{figure}[b!]
  \centering
{\includegraphics[width=.48\textwidth]{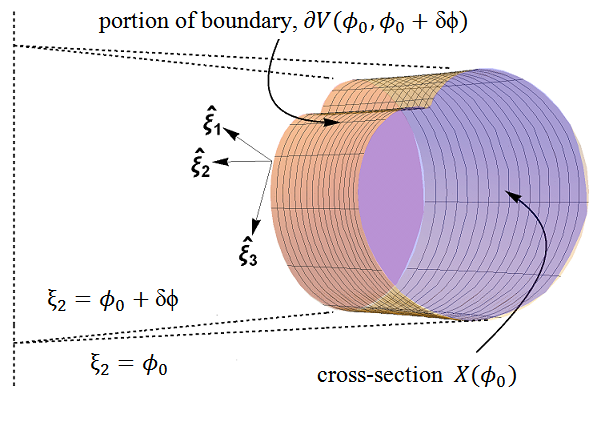}}
  \caption{Color) Volume $V'$ with cross-section $X$ bounded by the surfaces $\xi_2=\phi_0$  and $\phi_0+\delta\phi$. The faces of $V'$ are $\partial V(\phi_0,\phi_0 + \delta \phi), X(\phi_0)$ and $X(\phi_0+\delta \phi)$.}
  \label{fig: pillbox SM}
\end{figure}
Here we derive  the geometric integral formula:

\begin{eqnarray}
\bm{F}(\phi_0) \cdot \bm{\hat{\nu}}(\phi_0) &=& \alpha \iint\displaylimits_{\partial V(\phi_0,\phi_0+\delta\phi)} dS( \bm{\hat{N}} \cdot \bm{\hat{n}}) \>  (\bm{\hat{\nu}}   \cdot  \bm{\hat{n}} ) \nonumber \\
 &+ & \alpha \> \delta\phi \>\> \iint\displaylimits_{X(\phi_0)} dS( \bm{\hat{\nu}} \times \bm{\hat{z}} \cdot \bm{\hat{n}}) \>  (\bm{\hat{\nu}}   \cdot  \bm{\hat{n}} ) \nonumber \\
 \label{eqn: integrated force}
\end{eqnarray}

\noindent used in section IV of the main text \cite{Main} for a sample with symmetry and arbitrary smooth cross-section $X$, parametrised by general orthogonal curvilinear coordinates $\xi_{1,2,3}$ shown in Fig. \ref{fig: pillbox SM}. In the main text we make the replacements in notation $\bm{\hat{\xi_1}} \to \bm{\hat{N}}$, the normal to the bounding surface $\partial V$, $\bm{\hat{\xi_2}} \to \bm{\hat{\nu}}$, the direction of symmetry and $\bm{\hat{\xi_3}} \to \bm{\hat{\tau}} = \bm{\hat{N}} \times \bm{\hat{\nu}}$. The net active force $\bm{F}(\phi_0)$ acting on this volume is given by 
\begin{eqnarray}
\bm{F}(\phi_0)=\iiint_{V'}h_1h_2h_3d\xi_1d\xi_2d\xi_3 \bm{f}_a \,,
\label{FV}
\end{eqnarray}
where the geometrical  scale factors $h_{1,2,3}$ are the ratios of the infinitesimal distances to infinitesimal changes $d\xi_{1,2,3}$ in the curvilinear coordinates (and should not, of course, be confused with the components of the ``molecular fields" $\bm{h}$).
Applying the divergence theorem to 
the component of $\bm{F}(\phi_0)$   along the direction $\bm{\hat{\xi_2}}$ 
enables us to convert the volume integral in (\ref{FV}) into an integral over the surface $\partial V'$ of $V'$:
\begin{eqnarray}
\bm{F}(\phi_0) \cdot \bm{\hat{\xi_2}}(\phi_0) = \alpha \int_{\partial V'} dS (\bm{\hat{\xi_1}}\cdot \bm{\hat{n}})(\bm{\hat{\xi_2}}\cdot \bm{\hat{n}}).
\end{eqnarray}
To evaluate this surface integral, we note that the surface $\partial V'$ of $V'$ can be divided into three parts: the portion of the sample surface $\partial V(\phi_0,\phi_0 + \delta \phi)$ that borders $V'$, and the two cross-sectional  ``caps" $X(\phi_0)$ and $X(\phi_0+\delta\phi)$ (see Fig. \ref{fig: pillbox SM}). Doing so 
gives
 three surface integrals to evaluate, the  first of which is:
\begin{eqnarray}
I^{\alpha}_{\partial V(\phi_0,\phi_0 + \delta \phi)} &=& \alpha \bm{\hat{\xi_2}}(\phi_0) \cdot \int_{\phi_0}^{\phi_0+\delta\phi} d \xi_2 \int d\xi_3 h_2 h_3 n_1 n_i \bm{\hat{\xi_i}} \nonumber \,.
\end{eqnarray}
Using the facts that the element of surface area $dS= d \xi_2  d\xi_3 h_2 h_3$, $n_1=\bm{\hat{N}}\cdot \bm{\hat{n}}$, $\bm{\hat{n}}=n_i \bm{\hat{\xi_i}}$ and, in  the notation of the main text, $\bm{\hat{\xi_2}}(\phi_0) =\bm{\hat{\nu}}$, we obtain the first term on the right hand side of Eq. (\ref{eqn: integrated force}):
\begin{equation}
I^{\alpha}_{\partial V(\phi_0,\phi_0 + \delta \phi)} \approx  \alpha \int_{\partial V(\phi_0,\phi_0 + \delta \phi)} dS (\bm{\hat{N}}\cdot \bm{\hat{n}})(\bm{\hat{\nu}}\cdot \bm{\hat{n}}).
\label{sides}
\end{equation}
Now evaluating the integrals across the cross-sections, it is convenient to combine them as follows:
\begin{eqnarray}
I^{\alpha}_{X(\phi_0)}  &=& \alpha \bm{\hat{\xi_2}}(\phi_0) \cdot \int d\xi_1 d\xi_3 h_1 h_3 n_2 \left(-n_i \bm{\hat{\xi_i}}(\phi_0)\right) \nonumber \\
I^{\alpha}_{X(\phi_0+\delta\phi)} &=& \alpha \bm{\hat{\xi_2}}(\phi_0) \cdot \int d\xi_1 d\xi_3 h_1 h_3 n_2 \left(n_i \bm{\hat{\xi_i}}(\phi_0+\delta\phi)\right) \nonumber
\end{eqnarray}
\noindent so that
\begin{eqnarray}
I^{\alpha}_{X(\phi_0)} + I^{\alpha}_{X(\phi_0+\delta\phi)} &\approx & \alpha \delta \phi \bm{\hat{\xi_2}} \cdot \int d\xi_1 d\xi_3 h_1 h_3 n_2 n_i \partial_2 \bm{\hat{\xi_i}}(\phi_0) \nonumber \\
&\approx & \alpha \delta\phi \int d\xi_1 d\xi_3 h_1 h_3 n_2 \bm{\hat{\xi_2}}\cdot \left( n_i  \partial_2 \bm{\hat{\xi_i}}\right) \nonumber \\
\label{2nd term}
\end{eqnarray}
\noindent again taking $\bm{\hat{\xi_2}}$ inside the integral sign.
The second term $I^{\alpha}_{X(\phi_0)} + I^{\alpha}_{X(\phi_0+\delta\phi)}$ can be simplified by noting that $\bm{\hat{\xi_2}}\cdot \partial_2 \bm{\hat{\xi_i}}=\partial_2\bm{\hat{\xi_2}}
\cdot   \bm{\hat{\xi_i}}-\bm{\hat{\xi_i}}\cdot \partial_2 \bm{\hat{\xi_2}}$. Since $\bm{\hat{\xi_2}}\cdot  
\bm{\hat{\xi_i}}=\delta_{i2}$, which is independent of $\phi$, the first term vanishes. The 
argument of the integral in (\ref{2nd term}) can then be rewritten $ \bm{\hat{\xi_2}}\cdot 
\left( n_i  \partial_2 \bm{\hat{\xi_i}}\right)=- n_i\bm{\hat{\xi_i}}\cdot 
\left(  \partial_2 \bm{\hat{\xi_2}}\right)=-\bm{\hat{n}\cdot 
\left(  \partial_2 \bm{\hat{\xi_2}}\right)$}, where we've used the fact that $n_i\bm{\hat{\xi_i}}=\bm{\hat{n}}$ (this simply being the decomposition of $\bm{\hat{n}}$ along the local coordinate axes $\bm{\hat{\xi_i}}$). Now, using the fact that $\partial_2 \bm{\hat{\xi_2}}=\partial_\phi\bm{\hat{\phi}}=- \bm{\hat{r}}$, where $\bm{\hat{r}}$ is the unit vector in the radial direction from the axis of toroidal symmetry, we obtain 
\begin{eqnarray}
I^{\alpha}_{X(\phi_0)} + I^{\alpha}_{X(\phi_0+\delta\phi)} &\approx &  \alpha \> \delta\phi \>\> \iint\displaylimits_{X(\phi_0)} dS \left(\left(\bm{\hat{\nu}} \times \bm{\hat{z}}\right) \cdot \bm{\hat{n}}\right) \>  (\bm{\hat{\nu}}   \cdot  \bm{\hat{n}} )\nonumber\\
\label{2nd term2}
\end{eqnarray}
where we've used the fact that $\bm{\hat{r}}=\bm{\hat{\nu}} \times \bm{\hat{z}}$.
Adding this expression for the contribution of the cross sections $X(\phi_0)$ and $X(\phi_0+\delta\phi)$ to the net toroidal force to that of the boundary $\partial V$ as given by (\ref{sides}) immediately gives Eq. (16) of the main text.


\noindent \emph{High slenderness limit} In the case of torsional symmetry with an arbitrary (smooth) cross-section $X$ where the volume has a high slenderness, $\sigma$, the second term in Eq. (16) of the main text may be dropped if the first term is non-zero. To see this, suppose that the length-scale in the $\bm{\hat{\xi_1}}$- and $\bm{\hat{\xi_3}}$-directions is $L$, while in the $ \bm{\hat{\xi_2}}$-direction it has a length scale of $\sigma L$. A very slender sample will therefore have $\sigma \gg 1$, whereas a ``fat" sample will have $\sigma \approx 1$. The first term in Eq. (16) is proportional to $L^2 \sigma$, whereas the second term is proportional to $\L^2$ and so can be neglected compared with the first term.

\section{\uppercase{ Condition on the vorticity for flow in the ``frozen director" regime}}

Under the assumptions of section IV of \cite{Main} (i.e., torsional symmetry and that the pressure gradient is orthogonal to the direction of torsional symmetry $\bm{\hat{\nu}}$), we can provide a simple geometric statement of Eq. (\ref{eqn: simplified Navier Stokes}) in terms of the vorticity of the flow $\bm{\Omega} \equiv \nabla \times \bm{v}$ and the anchoring condition of the director at the boundary. If we apply the divergence theorem to Eq. (\ref{eqn: simplified Navier Stokes}) over the pillbox volume $V'$ in Fig. 1(a) of \cite{Main}, we obtain
\begin{equation}
 \eta \iint_{\partial V'} dS \bm{\hat{N}} \times \bm{\Omega} + \iint_{\partial V'} dS \bm{\hat{N}} P = \alpha \iint_{\partial V'} dS \bm{\hat{N}} \cdot \bm{\hat{n_0}} \> \bm{\hat{n_0}},
\end{equation}
\noindent using $\nabla^2 \bm{v} = -\nabla \times \Omega$ and the divergence theorem applied to $\nabla \times \Omega$. In the case of a slender sample, if we now again project along the direction of symmetry $\bm{\hat{\nu}}(\phi_0)$ and sum over all such pillbox volumes comprising the sample, we may take $\bm{\hat{\nu}}$ inside the integral and obtain
\begin{equation}
\iint_{\partial V} dS \> \bm{\hat{\tau}} \cdot \bm{\Omega} \approx -\frac{\alpha}{\eta} \int_{\partial V} dS \> ( \bm{\hat{N}} \cdot \bm{\hat{n_0}}) \>  (\bm{\hat{\nu}}   \cdot  \bm{\hat{n_0}} ),
\label{eqn: integrated force balance}
\end{equation}

\noindent where $\partial V$ is now the boundary of the sample and $\bm{\hat{\tau}}= \bm{\hat{N}} \times \bm{\hat{\nu}}$.

\section{\uppercase{Analysis of the "Frederiks cell" active pump}}
Our starting point is the following parametrisation for the director field (see Fig. 5(a) in the main text \cite{Main}):
\begin{equation}
\bm{\hat{n}} = \sin \theta (x) \> \bm{\hat{x}} + \cos \theta (x)  \bm{\hat{t}}(x) \,,
\label{eq: ansatz Frederiks}
\end{equation}
where we've defined an $x$-dependent unit vector orthogonal to  $\bm{\hat{x}}$:
\begin{equation}
 \bm{\hat{t}}(x) \equiv  \sin \zeta (x) \> \bm{\hat{y}} + \cos \zeta (x) \> \bm{\hat{z}}\,.
\label{eq: ansatz Frederiks2}
\end{equation}
\noindent The boundary conditions, expressed in terms of the angles $\theta (x)$ and $\zeta (x)$, are $\theta(\pm L/2) =0$ and $\zeta(\pm L/2)=\pm \zeta_0 /2$. Making the single Frank constant approximation ($K_{1,2,3} \equiv K$), the Euler-Lagrange equations for $\nh$ are:
\begin{equation}
K {d^2 \bm{\hat{n}}(x)\over dx^2} + g n_x(x) \bm{\hat{x}} = \mu (x) \bm{\hat{n}}
\label{ELGFR1}
\end{equation}
where $\mu (\bm{x})$ is the Lagrange multiplier ensuring that $\bm{n}^2 =1$ and $g \equiv \epsilon_0 \Delta \chi \> E^2$ can be experimentally controlled with an electric field $E$.

Since $\bm{\hat{t}}(x)$ is a unit vector in the plane orthogonal to $ \bm{\hat{x}}$, its derivative must be a vector orthogonal to itself in that plane, whose magnitude is $\zeta'(x)\equiv {d\zeta\over dx}$. This observation implies
\begin{equation}
{d \bm{\hat{t}}(x)\over dx} =\zeta'(x)  \bm{\hat{t}}\times\bm{\hat{x}}\,,
\label{ELGFR2}
\end{equation}
repeated application of which in turn leads to a natural decomposition of ${d^2 \bm{\hat{n}}(x)\over dx^2}$  along the orthonormal triad  
$ \bm{\hat{t}}$, $\bm{\hat{x}}$, and  $\bm{\hat{t}}\times\bm{\hat{x}}$:
\begin{eqnarray}
{d^2 \bm{\hat{n}}(x)\over dx^2}&=&[\theta''\cos \theta-\theta'^2 \sin \theta]  \bm{\hat{x}} \nonumber \\
&&-[\zeta'^2\cos \theta+\theta''\sin \theta+\theta'^2 \cos \theta]\bm{\hat{t}} \nonumber \\
&&+[\zeta''\cos \theta-2\theta'\zeta'\sin \theta]\bm{\hat{t}}\times\bm{\hat{x}}.
\label{ELGFR3}
\end{eqnarray}

Comparing this expression with the Euler-Lagrange Eq. (\ref{ELGFR1}), this must also equal:
\begin{eqnarray}
\frac{1}{K}[\mu(x) \bm{\hat{n}} - g n_x \bm{\hat{x}}] = \frac{\mu-g}{K}\sin \theta \> \bm{\hat{x}} + \frac{\mu}{K}\cos \theta \> \bm{\hat{t}}.
\label{ELGFR4}
\end{eqnarray}
Equating the $\bm{\hat{x}}$ components of these two expressions enables us to solve for the Lagrange multiplier $\mu(x)$:
\begin{eqnarray}
\mu(x)=K[\theta''\cot \theta-\theta'^2]+g\,.
\label{mu}
\end{eqnarray}
From the $\bm{\hat{t}}\times\bm{\hat{x}}$ components of (\ref{ELGFR3}) and (\ref{ELGFR4}) we immediately obtain
\begin{eqnarray}
\zeta''\cos\theta=2\theta'\zeta'\sin \theta\,,
\label{zeta1}
\end{eqnarray}
which can be reorganized as
\begin{eqnarray}
{\zeta''\over\zeta'}={2\theta'\sin \theta\over\cos\theta}\,.
\label{zeta2}
\end{eqnarray}
The left hand side of this expression is ${d\ln \zeta'\over dx}$, while the right hand side is $-2{d\ln \cos\theta\over dx}$. Hence this equation can be rewritten as 
\begin{eqnarray}
{d\over dx}\left(\ln(\zeta'\cos^2\theta)\right)=0\,,
\label{zeta3}
\end{eqnarray}
or, 
\begin{eqnarray}
\zeta' &=& C \sec^2 \theta
\label{zp}
\end{eqnarray}
where $C$ is a constant of integration which can be fixed by the boundary conditions, which, taken together with Eq. (\ref{zp}), imply $ \zeta_0 = C \int_{-L/2}^{L/2} \sec^2 \theta \> dx$.

From the $\bm{\hat{t}}$ component of (\ref{ELGFR4}), we obtain, using (\ref{mu}) and (\ref{zp}),
\begin{eqnarray}
-K[C^2\sec^3 \theta+\theta''\sin \theta+\theta'^2 \cos \theta]&=&\nonumber\\\left(K[\theta''\cot \theta-\theta'^2]+g\right)\cos\theta \,,
\label{theta1}
\end{eqnarray}
which can be solved for $\theta''$:
\begin{eqnarray}
\theta''=-C^2\sec^3 \theta\sin \theta-{g\over K} \sin\theta\cos \theta \,.
\label{theta2}
\end{eqnarray}
The first integral of this gives:

\begin{eqnarray}
\label{eq: Frederiks theta}
\theta'^2 &=& \frac{g}{K} \left(\sin^2 \theta_0 - \sin^2 \theta \right)+C^2 \left(\sec^2 \theta_0 - \sec^2 \theta \right) \nonumber \\
\end{eqnarray}
\noindent where $\theta_0$ is the maximum value taken by $\theta$. (\ref{eq: Frederiks theta}) can be solved analytically in closed form:
\begin{equation}
\frac{\sin \theta}{\sin \theta_0} = \mathrm{sn} \left[ \left(x +\frac{L}{2} \right) \sqrt{\frac{g}{K} + C^2 \sec^2 \theta_0} \>\> \Bigg| \>\> \frac{g \sin^2 \theta_0}{g + KC^2 \sec^2 \theta_0}  \right]
\end{equation}
\noindent where $\mathrm{sn} \left[z|m\right]$ is the Jacobi elliptic function with elliptic modulus $m$. It is more informative, however, to look at the $\theta_0 \ll 1$ limit of the derivative of (\ref{eq: Frederiks theta}), which is
\begin{equation}
\label{eq: Frederiks Duffing}
\theta'' + \theta \left(\frac{g}{K} + C^2 \right)-\frac{2}{3} \theta^3 \left( \frac{g}{K}-2C^2 \right) =0.
\end{equation}
\noindent This is recognisable as Duffing's equation; its solution can be expanded \cite{Bender1978} in the small parameter $\epsilon \equiv -\theta_0^2 (2g/K -4 C^2)/(3g/K+3C^2)$ as
\begin{equation}
\label{eq: Frederiks Duffing solution}
\frac{\theta\left[t(x)\right]}{\theta_0} = \cos t + \epsilon \left[\frac{1}{32} \left(\cos 3t - \cos t \right) - \frac{3}{8} t \> \sin t \right] + \mathcal{O} (\epsilon^2),
\end{equation}
\noindent where we have defined $t(x) \equiv x \> \sqrt{g/K + C^2}$ for reasons which we now explain. Duffing's equation can be used to approximate a planar pendulum for which the period is $T=2 \pi (1-3 \epsilon /8) + \mathcal{O}(\epsilon^2)$. In our case, in order to satisfy the boundary condition $\theta(\pm L/2) = 0$ (with the maximum attained at $x=0$) we require $t(L/2) = T/4$, or
\begin{equation}
\label{eq: Frederiks Duffing period}
\frac{L}{2} \sqrt{\frac{g}{K}+C^2} = \frac{\pi}{2} \left(1-\frac{3}{8} \epsilon \right).
\end{equation}
Close to the Frederiks transition, where the field is just above its critical value  $g_c$, $\theta(x) = \theta_0 \cos (\pi x/L) + \mathcal{O}(\epsilon)$, and to leading order the boundary condition on $\zeta$ is therefore $\zeta_0 = L C (1+\theta_0^2 /2)$. When $g=g_c$, the critical value of the field, Eq. (\ref{eq: Frederiks Duffing period}) is satisfied with $\theta_0=0$, in which case $C=\zeta_0/L$ and $\epsilon=0$, allowing us to conclude
\begin{equation}
 g_c ={ K\over L^2} (\pi^2-\zeta_0^2).
\end{equation}
Then writing $g=g_c + \Delta g$ and expanding Eq. (\ref{eq: Frederiks Duffing period}) to first order in the small quantities $\Delta g \> L^2/K$ and $\theta_0^2$, we can now relate the maximum amplitude $\theta_0$ to the incremental field $\Delta g$:
\begin{equation}
\theta_0^2 = \frac{2 \Delta g \> L^2}{K \left(\pi^2-\zeta_0^2 \right)}.
\end{equation}


\section{\uppercase{Analysis of the active nematic on a toroidal  shell}}

Here we provide the calculation for the optimal value of $\omega$ in the ansatz 

\begin{equation}
\bm{{\hat n}}=\frac{\omega \xi}{\xi + \cos \psi} \>\> \hat{\bm{\psi}}+\sqrt{1-\left(\frac{\omega \xi}{\xi + \cos \psi}\right)^2} \>\> \hat{\bm{\phi}},
\label{equation: ansatz}
\end{equation}

\noindent for a toroidal nematic shell. The Frank free energy
\begin{eqnarray}
F &=& \frac{1}{2}\int d^3 \bm{r} [K_{1} \left(\nabla\cdot\hat{\bm{n}}\right)^2+ K_{2}\left(\hat{\bm{n}}\cdot \left( \nabla \times \hat{\bm{n}} \right)\right)^2 \nonumber \\
&& \>\>\>\>\>\>\>\>\>\>\>\>\> + K_{3}\left|\hat{\bm{n}}\times \left( \nabla \times \hat{\bm{n}} \right)\right|^2 ] \nonumber \\ &&  - K_{24} \int d\bm{S} \cdot [\hat{\bm{n}} \nabla \cdot \hat{\bm{n}} + \hat{\bm{n}} \times (\nabla \times \hat{\bm{n}})] \nonumber \\
&\equiv & F_1 + F_2 + F_3 + F_{24},
\label{SM Frank FE}
\end{eqnarray}
\noindent must be minimized by varying the parameter $\omega$. For completeness, we include the saddle-splay contribution to the Frank free energy (the $K_{24}$ term) which can be relevant in curved geometries where the director field is not completely specified on the boundary, although we will show that there is no contribution in this case. For a thin shell, the $\rho$-integral runs from $1-\delta$ to $1$, whereas the surface integral is taken over the surfaces $\rho=1$ and $\rho=1-\delta$ where $\delta \ll 1$. In toroidal coordinates $(\rho,\phi,\psi)$, the scale factors are $(h_{\rho},h_{\phi},h_{\psi}) = R_2 (1,\xi+\rho \cos \psi,\rho)$; for the ansatz (\ref{equation: ansatz}) it follows that
\begin{eqnarray}
\nabla \cdot \hat{\bm{n}} &=& \frac{\omega \xi^2 (1-\rho) \sin \psi}{R_2 \rho (\xi + \rho \cos \psi)(\xi+\cos \psi)^2} \\
\label{eq: div n}
\nabla \times \hat{\bm{n}} &=& \hat{\bm{\rho}} \frac{\sin \psi [\rho + \frac{\omega^2 \xi^3 (1-\rho)}{(\xi+\cos \psi)^3}]}{R_2 \rho (\xi+\rho \cos \psi) \sqrt{1-\frac{\omega^2 \xi^2}{(\xi+\cos \psi)^2}}} \\
&-& \hat{\bm{\phi}} \frac{\omega \xi}{R_2 \rho (\xi+\cos \psi)} + \hat{\bm{\psi}} \frac{\cos \psi \sqrt{1-\frac{\omega^2 \xi^2}{(\xi+\cos \psi)^2}} }{R_2 (\xi+\rho \cos \psi)}.  \nonumber
\label{eq: curl n}
\end{eqnarray}
As mentioned above, our first observation is that there is no contribution from saddle-splay. To see this, note that for a shell of any thickness, $F_{24}$ comprises two surface integrals, on the outside and the inside of the shell. Plugging in the expressions in (\ref{eq: div n}) and (\ref{eq: curl n}) to (\ref{SM Frank FE}) over the outside surface only at $\rho=\rho_0$:
\begin{equation}
\frac{F_{24}^{\rm{outside}}}{K_{24}} = -R_2 \int d\phi d\psi \left(\rho_0 \cos \psi + \frac{\omega^2 \xi^3}{(\xi+\cos \psi)^2} \right).
\end{equation}

As the first term is zero, only the second term remains, and the integral is independent of $\rho_0$. Thus the integral over the inside surface exactly cancels that on the outside surface and $F_{24}=0$.

For the volume integrals, since $\nabla \cdot \hat{\bm{n}}=0$ at $\rho=1$, we also have $F_1=0$. Since we expect that  close to the transition  $\omega$ will be small, we expand $F_2$ and $F_3$ to $\mathcal{O}(\omega^4)$. The Frank free energy  then takes the form $A \omega^4 + B \omega^2 + C + \mathcal{O}(\omega^6)$, which has its minimum at $\omega^2 = -B/2A$, provided that $A>0$ and $B<0$ . It is sufficient to expand the coefficient of $\omega^2$ and $\omega^4$ in $1/\xi$ to leading order; doing this gives
\begin{eqnarray}
\frac{F_2}{K_2} &=& \frac{\delta}{2} \int d\phi d\psi \frac{\omega^2 \xi^4}{(\xi+\cos \psi)^4} \left(1-\frac{\omega^2 \xi^2}{(\xi+\cos\psi)^2} \right) \nonumber \\
&\approx & \pi \delta \xi \int d\phi \>\> (\omega^2 -\omega^4), \\
\frac{F_3}{K_3} &=& \frac{\delta}{2} \int d\phi d\psi \frac{\left(\cos\psi+\frac{\omega^2 \xi^3}{(\xi+\cos\psi)^2} \right)^2 + \frac{\sin^2 \psi}{1-\frac{\omega^2 \xi^2}{(\xi+\cos\psi)^2}}}{\xi + \cos \psi} \nonumber \\
&\approx & \pi \delta \xi \int d\phi \>\> (\rm{const} - \omega^2 \frac{5}{2\xi^2} + \omega^4).
\end{eqnarray}

For chiral symmetry breaking, since we require that $A>0$ and $B<0$, we see that $K_2/K_3$ must be at least of order $1/\xi^2$. Thus to leading order in $1/\xi$, we may neglect the higher order terms in $1/\xi$ of each coefficient and the $K_2$ contribution to the $\omega^4$ term, obtaining
\begin{equation}
\omega=\pm \sqrt{\frac{5}{4\xi^2} - \frac{K_2}{2 K_3}},
\end{equation}
\noindent provided that the expression under the square root is positive.


%